\documentclass[twocolumn, 10pt]{IEEEtran}
\usepackage{graphicx}
\usepackage{amssymb}
\usepackage{amsmath}
\usepackage{mathtools}
\usepackage{cite}
\usepackage{stfloats}
\usepackage{color}
\usepackage{subfigure}
\usepackage{caption}
\usepackage{dsfont}
\usepackage{psfrag}
\usepackage[mathscr]{euscript}
\usepackage{acronym}  % make an acronym
\usepackage{booktabs}
\usepackage{url}
\usepackage[table]{xcolor}

\acrodef{BS}{base station}
\acrodef{MEC}{mobile edge computing}
\acrodef{HetNet}{heterogeneous network}
\acrodef{VUE}{vehicular user equipment}
\acrodef{UE}{user equipment}
\acrodef{LoS}{Line-of-Sight}
\acrodef{NLoS}{Non-Line-of-Sight}
\acrodef{UDN}{Ultra-Dense Network}

\acrodef{PPP}{Poisson point process}
\acrodefplural{PPP}[PPPs]{Poisson point processes}

\acrodef{PDF}{probability density function}
\acrodef{CDF}{cumulative distribution function}
\acrodef{CCDF}{complementary cumulative distribution function}
\acrodef{SIR}{signal-to-interference ratio}
\acrodef{SINR}{signal-to-interference-plus-noise ratio}
\acrodef{SNR}{signal-to-noise ratio}
\acrodef{PGFL}{probability generating functional}
\acrodef{ASE}{area spectral efficiency}

\acrodef{V2X}{vehicle-to-everything}
\acrodef{RGM}{random Gauss-Markov}
\acrodef{STP}{successful transmission probability}
\acrodef{RSU}{roadside unit}
\acrodef{VUE}{vehicular user equipment}

\acrodef{P-K}{Pollaczek-Khinchin}
\acrodef{KKT}{Karush-Kuhn-Tucker}

\acrodef{MISO}{multi-input single-output}
\acrodef{MRT}{maximal ratio transmission}

\acrodef{STP}{successful transmission probability}

\acrodef{D2D}{device-to-device}
\acrodef{V2V}{vehicle-to-vehicle}

\acrodef{BEF}{best effort first}
\acrodef{BEL}{best effort last}

\acrodef{i.i.d.}{independently and identically distributed}

\acrodef{NOMA}{non-orthogonal multiple access}

\acrodef{CSIT}{channel state information at transmitters}
\acrodef{AWGN}{additive white Gaussian noise}

\acrodef{IoT}{Internet of Things}
\acrodef{AI}{artifical intelligence}
\newtheorem{theorem}{Theorem}
\newtheorem{lemma}{Lemma}
\newtheorem{corollary}{Corollary}

\newtheorem{remark}{Remark}

\newtheorem{problem}{Problem}

\newcommand{\Prob}[1]{\mathbb{P}\left[#1\right]}

\newcommand{\Expt}[1]{\mathbb{E}\left[#1\right]}

\newcommand{\CP}{\text{cp}}
\newcommand{\CM}{\text{cm}}
\newcommand{\SNR}{\text{SNR}}

\newcommand{\velConst}{v}

\newcommand{\dataOut}{D}

\newcommand{\targetTot}{\theta}
\newcommand{\targetK}[1]{\targetTot_{#1}}

\newcommand{\BStotal}{K}
\newcommand{\BSset}{\mathcal{K}}

\newcommand{\BScovK}[1]{R_{#1}}

\newcommand{\tCommVarK}[1]{t_{\CM,#1}}

\newcommand{\linDisMaxK}[1]{\ell_{#1}}
\newcommand{\PathLoss}[1]{g_{#1}}

\newcommand{\CH}[1]{\mathbf{h}_{#1}}
\newcommand{\CHtra}[1]{\mathbf{h}^{\text{H}}_{#1}}
\newcommand{\PL}{\alpha}
\newcommand{\noise}{N_o}
\newcommand{\noiseVar}{{z}}

\newcommand{\CPUunit}{C}

\newcommand{\CPUenergy}[1]{\varphi_{#1}}

\newcommand{\enerTot}{E_{\text{tot}}}
\newcommand{\enerTotTild}{\widetilde{E}_{\text{tot}}}

\newcommand{\enerK}[1]{E_{#1}}
\newcommand{\enerKTild}[1]{\widetilde{E}_{#1}}

\newcommand{\enerOpt}{\enerTot^{*}}

\newcommand{\BSpowerOptTmpK}[1]{\BSpowerK{k}^{\dagger}}

\newcommand{\BSpowerOptDDagK}[1]{\BSpowerK{k}^{\ddagger}}

\newcommand{\BSpowerKopt}[1]{\BSpowerK{#1}^*}

\newcommand{\BScovReq}[1]{R_{\text{req},#1}}
\newcommand{\BScovReqSingle}{R_{\text{req}}}

\newcommand{\BSpowerKoptSet}{\BSpowerSet{}^*}

\newcommand{\BW}{B}

\newcommand{\BSfreqM}[1]{F_{#1}}
\newcommand{\indexFreq}{f}
\newcommand{\BSfreqK}[1]{\indexFreq_{#1}}
\newcommand{\BSfreqSet}[1]{\mathbf{\indexFreq}_{#1}}
\newcommand{\BSfreqOptSet}{\mathbf{\indexFreq}^*}

\newcommand{\BSfreqOptK}[1]{\indexFreq^{*}_{#1}}

\newcommand{\BSpowerM}[1]{P_{#1}}
\newcommand{\indexPower}{p}
\newcommand{\BSpowerK}[1]{\indexPower_{#1}}
\newcommand{\BSpowerSet}[1]{\mathbf{\indexPower}_{#1}}
\newcommand{\BSpowerOptSet}{\mathbf{\indexPower}^*}
\newcommand{\BSpowerOptK}[1]{\indexPower^{*}_{#1}}

\newcommand{\BSantenna}{M}
\newcommand{\recvPower}[1]{{y}_{#1}}
\newcommand{\MRTweight}[1]{\mathbf{w}_{#1}}
\newcommand{\MRTsignal}[1]{s_{#1}}

\newcommand{\achievRate}[1]{C_{#1}}

\newcommand{\MRTFunc}[1]{G\left(#1\right)}
\newcommand{\MRTFuncInv}[1]{G^{-1}\left(#1\right)}

\newcommand{\userSet}{\mathcal{U}}
\newcommand{\userTot}{U}

\newcommand{\userOrder}[1]{a_{#1}}
\newcommand{\userOrderIndexSet}{\mathcal{O}}

\newcommand{\velK}[1]{\velConst_{#1}}

\newcommand{\workload}[1]{C_{#1}}
\newcommand{\result}[1]{D_{#1}}

\newcommand{\taskSplit}[1]{x_{#1}}
\newcommand{\taskSplitMax}[1]{\taskSplit{#1}^{(\text{m})}}
\newcommand{\taskSplitSet}[1]{\mathbf{x}_{#1}}
\newcommand{\taskSplitOptSet}{\taskSplitSet{}^{*}}
\newcommand{\taskSplitOptK}[1]{\taskSplit{#1}^{*}}
\newcommand{\taskSplitOptTmpSet}{\taskSplitSet{}^{\dagger}}

\newcommand{\taskSplitOptDDagSet}{\taskSplitSet{}^{\ddagger}}
\newcommand{\taskSplitOptDDagK}[1]{\taskSplit{#1}^{\ddagger}}

\newcommand{\compResultK}[1]{\dataOut\taskSplit{#1}}

\newcommand{\tCompK}[1]{t_{\CP,#1}}
\newcommand{\tCompSet}{\mathbf{t}_{\CP}}
\newcommand{\tCompOptSet}{{\mathbf{t}}^{\dagger}_{\CP}}

\newcommand{\tStartCP}[1]{s_{\CP,#1}}
\newcommand{\tStartCPSet}[1]{\mathbf{s}_{\CP{#1}}}
\newcommand{\tStartCPOptSet}{\mathbf{s}^{*}_{\CP}}
\newcommand{\tStartCPOptTmpSet}{\mathbf{s}^{\dagger}_{\CP}}
\newcommand{\tStartCPOptK}[1]{s^{*}_{\CP,#1}}

\newcommand{\eCompCap}[1]{\kappa_{#1}}
\newcommand{\eCompConst}[1]{\varphi_{#1}}

\newcommand{\tCommK}[1]{t_{\CM,#1}}
\newcommand{\tCommSet}[1]{\mathbf{t}_{\CM{#1}}}
\newcommand{\tCommOptSet}{\tCommSet{}^*}
\newcommand{\tCommOptK}[1]{t^{*}_{\CM,#1}}
\newcommand{\tCommOptTmpSet}{\mathbf{t}^{\dagger}_{\CM}}

\newcommand{\tStartCM}[1]{s_{\CM,#1}}
\newcommand{\tStartCMSet}{\mathbf{s}_{\CM}}
\newcommand{\tStartCMOptSet}{\mathbf{s}^{*}_{\CM}}
\newcommand{\tStartCMOptTmpSet}{\mathbf{s}^{\dagger}_{\CM}}
\newcommand{\tStartCMOptK}[1]{s^{*}_{\CM,#1}}

\newcommand{\batchSet}{\bar{\userSet}}

\newcommand{\onlineEnerTot}{E_{\text{online}}}
\newcommand{\onlineEnerOpt}{\onlineEnerTot^{*}}

\newcommand{\perspecCP}[1]{c^{(\CP)}_{#1}}

\newcommand{\KKTFunc}[2]{H_{#1}\left(#2\right)}
\newcommand{\KKTFuncInv}[2]{H_{#1}^{-1}\left(#2\right)}
\newcommand{\KKTFuncTild}[2]{\tilde{H}_{#1}\left(#2\right)}
\newcommand{\KKTFuncTildInv}[2]{\tilde{H}_{#1}^{-1}\left(#2\right)}

\newcommand{\KKTeq}[1]{\gamma_{#1}}
\newcommand{\KKTeqOpt}[1]{\KKTeq{#1}^*}

\newcommand{\CPUunitK}[1]{\CPUunit_{#1}}

\newcommand{\tArrivK}[1]{T^{(\text{a})}_{#1}}
\newcommand{\tDeparK}[1]{T^{(\text{d})}_{#1}}

\newcommand{\userBarSet}{\bar{\mathcal{U}}}
\newcommand{\userBarTot}{\bar{U}}
\newcommand{\userOrderBar}[1]{\bar{a}_{#1}}
\newcommand{\velBar}[1]{\bar{v}_{#1}}
\newcommand{\idxBar}[1]{\bar{#1}}
\newcommand{\workloadBar}[1]{\bar{C}_{#1}}
\newcommand{\resultBar}[1]{\bar{D}_{#1}}
\newcommand{\taskSplitBar}[1]{\bar{x}_{#1}}
\newcommand{\tArrivBar}[1]{\bar{T}^{(\text{a})}_{#1}}
\newcommand{\tDeparBar}[1]{\bar{T}^{(\text{d})}_{#1}}
\newcommand{\BSpowerBar}[1]{\bar{p}_{#1}}
\newcommand{\tCommBar}[1]{\bar{t}_{\text{cm},#1}}
\newcommand{\BSfreqBar}[1]{\bar{f}_{#1}}
\newcommand{\BSfreqBarSet}{\bar{\mathbf{f}}}
\newcommand{\tStartCPBar}[1]{\bar{s}_{\text{cp},#1}}
\newcommand{\tStartCPBarSet}{\bar{\mathbf{s}}_{\text{cp}}}
\newcommand{\enerTotBar}{\bar{E}_{\text{tot}}}

\newcommand{\taskSplitOptSingle}{\taskSplitSet{}^{\star}}
\newcommand{\BSfreqOptSingle}{{\mathbf{f}}^{\star}}
\newcommand{\BSpowerOptSingle}{{\mathbf{p}}^{\star}}

\newcommand{\tCommOptSingle}{{\mathbf{t}}^{\star}_{\CM}}

\acrodef{BS}{base station}
\acrodef{AP}{access point}
\acrodef{HD}{half-duplex}
\acrodef{FD}{full-duplex}
\acrodef{IC}{interference cancellation}
\acrodef{HDHN}{hybrid-duplex heterogeneous network}
\acrodef{TDD}{time-division duplexing}
\acrodef{FDD}{frequency-division duplexing}
\IEEEoverridecommandlockouts % Make sure to have this to present ack

\begin{document}

%---------------------------------------------------------------------------%
%                     title, title footnote, header                         %
%---------------------------------------------------------------------------%

\title{Energy-efficient Cooperative Offloading for Edge Computing-enabled Vehicular Networks}

\author{
\IEEEauthorblockN{
        Hewon Cho,
        % \textit{Student Member, IEEE},
        Ying Cui,
         \textit{Member, IEEE}, 
        and 
        Jemin Lee
         \textit{Member, IEEE}
}
	\thanks{
	The material in this paper was presented, in part, at the IEEE International
	Conference on Communications, Dublin, Ireland, June 2020 \cite{ChoCuiLee:20}.
}
	\thanks{
	H.\ Cho is with
	Daegu Gyeongbuk Institute of Science and Technology (DGIST),
	333 Techno Jungang-daero, Republic of Korea 42988
	(e-mail: {nb00040@dgist.ac.kr}, {jmnlee@dgist.ac.kr}).
}
\thanks{
	Y.\ Cui is with the Department of Electronic Engineering, Shanghai Jiao Tong University, Shanghai
	200240, China (e-mail: cuiying@sjtu.edu.cn).
}
\thanks{
	J. Lee was with the Department of Information and Communication Engineering (ICE), DGIST, Daegu 42998, Republic of Korea. She is now with the Department of Electrical and Computer Engineering,
	Sungkyunkwan University (SKKU), Suwon 16419, Republic of Korea (e-mail:
	jemin.lee@skku.edu).
}
}

\maketitle %% make the title area

%
%\markboth{Submitted to IEEE Journal on Selected Areas in Communications}{\title}

%
%%%%%%%%% uncomment this section for a 2-column formt %%%%%%%
%%%%%%%%% [begin] %%%%%%%%
%\thispagestyle{empty}
%\textcolor{blue}{\framebox{\textsf{Today: \today}}}\\
\newcommand{\red}[1]{{\textcolor[rgb]{1,0,0}{#1}}}
\newcommand{\blue}[1]{{\textcolor[rgb]{0,0,1}{#1}}}
\newcommand{\green}[1]{{\textcolor[rgb]{0.184,0.616,0.153}{#1}}}
\newcommand{\JL}[1]{{\textcolor[rgb]{0,0,1}{[JL: #1]}}}
\newcommand{\HW}[1]{{\textcolor[rgb]{0,0,1}{[HW: #1]}}}
\newcommand{\change}{\color{magenta}}

\acresetall

%%---------------------------------------------------------------------------%
%%                           abstract and key words                          %
%%---------------------------------------------------------------------------%
\begin{abstract}
Edge computing technology has great potential to improve various computation-intensive applications in vehicular networks by providing sufficient computation resources for vehicles.
However, it is still a challenge to fully unleash the potential of edge computing in edge computing-enabled vehicular networks.
In this paper, we develop the energy-efficient cooperative offloading scheme for edge computing-enabled vehicular networks, which splits the task into multiple subtasks and offloads them to different \acp{RSU} located ahead along the route of the vehicle.
We first establish novel cooperative offloading models for the offline and online scenarios in edge computing-enabled vehicular networks.
In each offloading scenario, we formulate the total energy minimization with respect to the task splitting ratio, computation resource, and communication resource.
In the offline scenario, we equivalently transform the original problem to a convex problem and obtain optimal solutions for multi-vehicle case and single-vehicle case, respectively.
Furthermore, we show that the method proposed for the offline scenario can also be applied to solve the optimization problem in the online scenario.
Finally, through numerical results, by analyzing the impact of network parameters on the total energy consumption, we verify that our proposed solution consumes lower energy than baseline schemes.
\end{abstract}


%\clearpage
\acresetall
%%%%%%%%%%%%%%%%%%%%%%%%%%%%%%%%%%%%

%%---------------------------------------------------------------------------%
%%                           Sec: Introduction                               %
%%---------------------------------------------------------------------------%

\section{Introduction}

Due to the rapid development of vehicular networks, various computation-intensive applications, including road safety applications, traffic efficiency applications, infotainment applications, etc., are emerging.
Implementing such applications requires enormous computing resources for data processing \cite{LiMaGonZhoChe:21,LiuWanZenWanSonQiu:17,HeCheGeGui:16,NkeNkeIslKerrWadAla:19}.
However, the computing resource of a vehicle is usually limited,
making it hard for the vehicle itself to complete a computation task by the service completion deadline.
The edge computing technology, which enables computation at the network edge, has been regarded as a promising solution for tackling this issue \cite{DinLeeNiyWan:13,ETSIMEC:14,MaoYouZhanHuaLet:17,HasYauWu:19,LiuLiuTanYuWanShi:19,ParLee:20}.

There have been extensive works on an optimal offloading design for static users in edge computing-enabled networks \cite{LiaLiuLokHua:19,WanYaoZheSunSon:18,GuoSonCuiLiuJi:17,WanXuWanCui:18,BiYuaDuaZhoAbu:21,LiXuZhoLuHan:20,CheTenSunLiuWan:18,DinTanLaQue:17,WanSheWanWanLi:16,GuWanLiuFanFanZho:18}.
In this scenario, a static user offloads its task to one edge computing server for non-cooperative computation \cite{LiaLiuLokHua:19,WanYaoZheSunSon:18,GuoSonCuiLiuJi:17,WanXuWanCui:18,BiYuaDuaZhoAbu:21,LiXuZhoLuHan:20,CheTenSunLiuWan:18} or splits its task into subtasks and offloads them to multiple edge computing servers for cooperative computation \cite{DinTanLaQue:17,WanSheWanWanLi:16,GuWanLiuFanFanZho:18}.
Specifically, \cite{LiaLiuLokHua:19,WanYaoZheSunSon:18,GuoSonCuiLiuJi:17,WanXuWanCui:18,BiYuaDuaZhoAbu:21,LiXuZhoLuHan:20,CheTenSunLiuWan:18} consider non-cooperative offloading and optimize the offloading scheduling and communication and computation resource allocation during the offloading process to maximize the weighted sum of the offloading rates \cite{LiaLiuLokHua:19},
minimize the total latency for computing and transmitting a task \cite{WanYaoZheSunSon:18},
or minimize the total energy consumption for local computing and offloading \cite{GuoSonCuiLiuJi:17,WanXuWanCui:18,BiYuaDuaZhoAbu:21,LiXuZhoLuHan:20,CheTenSunLiuWan:18}.
On the other hand, \cite{DinTanLaQue:17,WanSheWanWanLi:16,GuWanLiuFanFanZho:18} study cooperative offloading and optimize the task splitting and communication and computation resource allocation during the offloading process to minimize the total latency for offloading \cite{DinTanLaQue:17} 
or minimize the total energy consumption for local computing and offloading \cite{WanSheWanWanLi:16,GuWanLiuFanFanZho:18}.
Due to the page limitation, we refer interested readers to \cite{LinZeaCheLabWan:20} for a recent survey on optimal offloading designs for static users in edge computing-enabled networks.
Note that the distance between a user and an edge computing server remains constant during an offloading process in the static scenario.
Without capturing the impact of user mobility on an offloading process,
the proposed solutions for the static scenario in \cite{LiaLiuLokHua:19,WanYaoZheSunSon:18,GuoSonCuiLiuJi:17,WanXuWanCui:18,BiYuaDuaZhoAbu:21,LiXuZhoLuHan:20,CheTenSunLiuWan:18,DinTanLaQue:17,WanSheWanWanLi:16,GuWanLiuFanFanZho:18} may no longer be apply to vehicles that move most of the time.

The optimal offloading design of edge computing-enabled vehicular networks has recently received increasing attention \cite{ZhaLiGonZha:19,WanFenZhaTanQue:19,DaiXuMahZha:19,SunGuZheDonValQin:20,SalLiuJanLiJia:20,ZhaGuoLiuZha:20,ZhuYanHuGaoSch:20,BozTarCor:18}.
For example, by optimizing the offloading scheduling and communication and computation resource allocation during the offloading process, \cite{ZhaLiGonZha:19,WanFenZhaTanQue:19,DaiXuMahZha:19,SunGuZheDonValQin:20} maximize the system utility,
\cite{SalLiuJanLiJia:20,ZhaGuoLiuZha:20} minimize the total latency,
and \cite{ZhuYanHuGaoSch:20} minimizes the total energy consumption.
Most of the works, including \cite{ZhaLiGonZha:19,WanFenZhaTanQue:19,DaiXuMahZha:19,SunGuZheDonValQin:20,SalLiuJanLiJia:20,ZhaGuoLiuZha:20,ZhuYanHuGaoSch:20}, offload a task to only one \ac{RSU} (i.e., edge computing server).
However, in a practical vehicular network, an \ac{RSU}'s coverage diameter is around 300m, and a vehicle moves at 20km/h-100km/h.
Thus, a vehicle may have a short connection time, e.g., 10-50 seconds, to a single \ac{RSU} \cite{LiuChePeiMahZha:20}.
If the vehicle offloads its task to an \ac{RSU} located far ahead of itself in advance, the computation energy consumption may reduce given sufficient computation time.
However, the communication energy consumption for downloading the computation result may still be considerable due to the short connection time to one \ac{RSU}.
Therefore, when the sizes of the workload and computation result are large and the velocity of a vehicle is high, offloading a task to a single \ac{RSU} (i.e., non-cooperative offloading) is obviously not energy-efficient or even not feasible.
To tackle the issue caused by the short connection time to each \ac{RSU}, \cite{BozTarCor:18} considers cooperative offloading in edge computing-enabled vehicular networks.
Specifically, \cite{BozTarCor:18} splits a vehicle's task into multiple subtasks and offloads them to different edge computing servers for cooperative computation.
Furthermore, \cite{BozTarCor:18} optimizes the task splitting to minimize the latency.
However, the resulting performance may not be satisfactory without optimizing the computation and communication resource allocation during the cooperative offloading process.
As opposed to cooperative offloading in the static scenario \cite{GuoSonCuiLiuJi:17,LiaLiuLokHua:19,WanYaoZheSunSon:18,WanXuWanCui:18,BiYuaDuaZhoAbu:21,LiXuZhoLuHan:20,CheTenSunLiuWan:18,WanSheWanWanLi:16,DinTanLaQue:17,GuWanLiuFanFanZho:18}, joint optimization of task splitting and resource allocation for cooperative offloading in edge computing-enabled vehicular networks is still open.
The challenge is caused by the dependency of the times that each vehicle arrives at and departs from the coverage areas of all \acp{RSU} on their arrival times to a vehicular network and their velocities.

To fully unleash the potential of edge computing in edge computing-enabled vehicular networks, we cooperatively utilize the communication and computation resources across \acp{RSU} and jointly optimize the task splitting and communication and computation resource allocation to minimize the total energy consumption at all \acp{RSU}.
In particular, we investigate two scenarios, i.e., the offline scenario and the online scenario.
The information about all vehicles that will enter the network is available in advance in the offline scenario but is unknown in advance in the online scenario.
The main contributions of this work can be summarized as below.
\begin{itemize}
	\item We establish novel cooperative offloading models for the offline and online scenarios in edge computing-enabled vehicular network.
	In each scenario, the task of every moving vehicle is split into multiple subtasks, which are offloaded to different \acp{RSU} located ahead along the route of the vehicle.
	Each \ac{RSU} completes the execution of each subtask before the corresponding vehicle enters its coverage area and finishes the transmission of the computation result to the vehicle when it is in the \ac{RSU}'s coverage area.
	Besides, the constraints reflecting the computation and communication operation orders for the subtasks of all moving vehicles are specified in terms of the CPU frequency, transmission power and time, computation starting time, and transmission starting time for each subtask at each \ac{RSU}.
	The models are much more complex than the one in the static scenario as the subtasks of each vehicle are offloaded to multiple \acp{RSU} and multiple vehicles pass through the coverage area of each \ac{RSU} at distinct and possibly overlapping intervals.
	To the best of our knowledge, this is the first work providing cooperative offloading models for moving vehicles that enable optimal cooperative offloading design for edge computing-enabled vehicular networks.
	\item In the offline scenario, we consider two cases, i.e., the multi-vehicle case and the single-vehicle case.
	In each case, we formulate the total energy minimization with respect to the task splitting ratio, computation resource, and communication resource.
	It is a challenging non-convex problem with considerably more variables and constraints than in cooperative offloading for the static scenario.
	To reslove the challenge, in each case, we first characterize an optimality property. Then, based on it, we equivalently transform the original non-convex problem to a convex problem with fewer variables and constraints.
	In the multi-vehicle case, we solve the equivalent convex problem with standard convex optimization algorithms.
	In the single-vehicle case, we further decompose the equivalent convex problem into several subproblems of much smaller sizes controlled by a master problem and obtain the closed-form optimal solutions for the subproblems and semi-closed-form optimal solution for the master problem.
	\item In the online scenario, we formulate the total energy minimization with respect to the task splitting ratio, the computation resource, and the communication resource of both new arrivals and leftovers (i.e., vehicles that already exist in the network when new arrivals enter the network).
	Here, leftovers represent the vehicles that already exist in the network when new arrivals enter the network.
	The optimization problem for the online scenario is even more challenging than for the offline scenario due to consecutive new arrivals to the vehicular network.
	We show that the underlying problem structure for the online scenario is similar to the one for the multi-vehicle case in the offline scenario andcan be solved using the same method.
	\item Finally, by numerical simulations, we analyze the impacts of vehicles' velocities and computation result sizes on the total energy consumption in the single-vehicle case and the multi-vehicle case, respectively.
	Furthermore, we show that the proposed solutions achieve significant gains in the total energy consumption over all baseline schemes.
	
\end{itemize}

The remainder of this paper is organized as follows.
Section \ref{sec:system_model} describes the system model.
Section \ref{sec:problem_formulation} formulates the energy minimization problem in the offline scenario and proposes low complexity optimal solutions for both the multi-vehicle case and single-vehicle case.
We then formulate the energy minimization problem in the online scenario and present its solution method in Section \ref{sec:online_case}.
Numerical results are provided in Section \ref{sec:numerical}.
Finally, conclusions are given in Section \ref{sec:conclusion}.

\emph{Notation:} The notation used throughout the paper is reported in Table \ref{table:notation}.

\begin{table}
	\caption{Notations used throughout the paper.} \label{table:notation}
	\begin{center}
		\rowcolors{2}%{green!10!yellow}{}
		{cyan!15!}{}
		\renewcommand{\arraystretch}{1.0}
		\begin{tabular}{ c  p{6.5cm} }
			\hline 
			{\bf Notation} & {\hspace{2.5cm}}{\bf Definition}
			\\
			\midrule
			\hline
			$\userOrder{k,m}$	&	Index of the vehicle that is $m$-th one entering \ac{RSU} $k$'s coverage area\\ \addlinespace
			$\workload{u}$		&	Size of workload of vehicle $u$'s task\\ \addlinespace
			$\result{u}$		&	Size of computation result of vehicle $u$'s task\\ \addlinespace
			$\BSfreqM{k}$		&	Maximum computing capability of \ac{RSU} $k$\\ \addlinespace
			$\BSfreqK{k,u}$; $\BSfreqSet{}$			&	CPU frequency used for executing vehicle $u$'s subtask $k$ at \ac{RSU} $k$; CPU frequency allocation\\ \addlinespace	
			$\BSpowerM{k}$		&	Maximum transmission power of \ac{RSU} $k$\\ \addlinespace
			$\BSpowerK{k,u}$; $\BSpowerSet{}$		&	transmission power used for transmitting vehicle $u$'s subtask $k$ at \ac{RSU} $k$; Transmission power allocation\\ \addlinespace					
			$\BScovK{k}$		&	Length of the interval of the road covered by \ac{RSU} $k$\\ \addlinespace
			$\BScovReq{u}$		&	Distance from \ac{RSU} $1$ to the point where vehicle $u$ requested the task\\ \addlinespace
			$\velK{u}$			&	Velocity of vehicle $u$\\ \addlinespace
			$\tStartCP{k,u}$; $\tStartCPSet{}$	&	Computation starting time of vehicle $u$'s subtask $k$ at \ac{RSU} $k$; computation starting time allocation\\ \addlinespace
			$\tStartCM{k,u}$; $\tStartCMSet$		&	Transmission starting time of vehicle $u$'s subtask $k$ at \ac{RSU} $k$; transmission starting time allocation\\ \addlinespace
			$\tArrivK{k,u}$; $\tDeparK{k,u}$	&	Arrival time of vehicle $u$ at \ac{RSU} $k$; departure time of vehicle $u$ at \ac{RSU} $k$\\ \addlinespace
			$\taskSplit{k,u}$; $\taskSplitSet{}$	&	Fraction of task $u$ that is executed at \ac{RSU} $k$; task splitting factor\\ \addlinespace			
			$\tCompK{k,u}$; $\tCompSet$	&	Computation time of vehicle $u$'s subtask $k$ at \ac{RSU} $k$; computation time allocation\\ \addlinespace
			$\tCommK{k,u}$; $\tCommSet{}$	&	Transmission time of vehicle $u$'s subtask $k$ at \ac{RSU} $k$; transmission time allocation\\ \addlinespace
			$\enerK{\CP,k,u}$		&	Computation energy consumption for executing vehicle $u$'s subtask $k$ at \ac{RSU} $k$\\ \addlinespace
			$\enerK{\CM,k,u}$		&	Communication energy consumption for transmitting the computation result from \ac{RSU} $k$ to vehicle $u$\\ \addlinespace
			$\enerTot$		&	Total energy consumption\\ \addlinespace
		\hline
		\end{tabular}
	\end{center}
\end{table}%

%%---------------------------------------------------------------------------%
%%                           Sec: System Model                               %
%%---------------------------------------------------------------------------%

\section{System Model in Offline Scenario}
\label{sec:system_model}

In this section, we present the network model and cooperative offloading model for the offline scenario of an edge computing-enabled vehicular network.
In the offline scenario, information about all vehicles that will enter the network during a certain period is available in advance.
Then, in Section \ref{sec:online_system}, we will present the network model and cooperative offloading model for the online scenario where the information about vehicles entering the network is updated from time to time.

\subsection{Network Model}

\begin{figure}[t!]
	\begin{center}
		{
			\includegraphics[width=1.0\columnwidth]{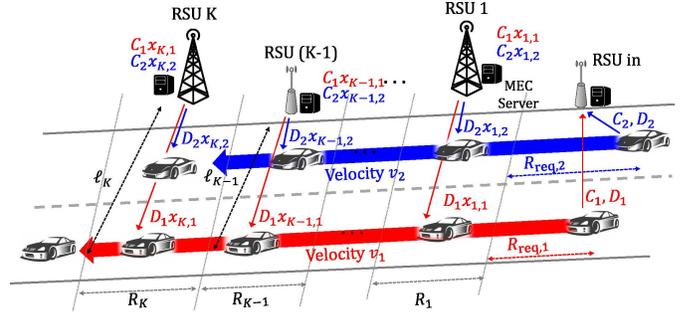}
		}
	\end{center}
	\caption{
		Illustration of the offline scenario.
	}
	\label{fig:system_model}
\end{figure}

We consider an edge computing-enabled vehicular network which consists of $\BStotal$ multi-antenna \acp{RSU}, denoted by $\BSset\triangleq\{1,\cdots,\BStotal\}$, located along a unidirectional road, as shown in Fig.\ref{fig:system_model}.\footnote{
	Note that the number of \acp{RSU}, $\BStotal$, is determined by the delay requirement of the tasks requested by the vehicles.
}
In the offline scenario, we consider $\userTot$ single-antenna vehicles, denoted by $\userSet\triangleq\left\{1,\cdots,\userTot\right\}$.
At time 0, the $\userTot$ vehicles enter the edge computing-enabled network and pass \acp{RSU} $1$, $2$, $\cdots$, $\BStotal$ successively.
Each \ac{RSU} is equipped with an edge computing server.
Thus, each \ac{RSU} has both communication capability and computation capability.
In reality, the road does not always pass through the center of each \ac{RSU}'s coverage area.
As a result, for each \ac{RSU} $k\in\BSset$, we introduce two measures, i.e., the length of the interval of the road covered by \ac{RSU} $k$, denoted by $\BScovK{k}$, and the maximum link length within \ac{RSU} $k$'s coverage interval, denoted by $\linDisMaxK{k}$.
For tractability, we assume that the $\BStotal$ road intervals covered by the $\BStotal$ \acp{RSU} form a partition of the road.

Each vehicle $u\in\userSet$ moves at a constant velocity, denoted by $\velK{u}$ (in m/s), along the road, and has a computation-intensive task, called task $u$ \cite{TanZhaRafQiDouNi:19}.
Due to limited local computation capability, each vehicle offloads its task and requests to obtain the result of its task before it leaves the edge computing-enabled vehicular network.\footnote{
	Our work can be extended by considering local computing at each vehicle.
}
Assume that at time 0, each vehicle has sent the input data of its task to the edge computing-enabled network.
Thus, we characterize a task by two parameters, i.e., the size of workload $\workload{u}>0$ (in number of CPU-cycles) and the size of the computation result $\result{u}>0$ (in bits).

Let $\BScovReq{u}$ (in meters) denote the distance to \ac{RSU} 1 at time 0.
The arrival time and the departure time of vehicle $u$ at \ac{RSU} $k$ are given as follows:
%%%
%%%
\begin{align}
	\tArrivK{k,u}
	=
	\frac{1}{\velK{u}}
	\left(
	\BScovReq{u}
	+
	\sum_{i=1}^{k-1}
	\BScovK{i}
	\right),
		\quad
		k\in\BSset
		,\,\,
		u\in\userSet,
\end{align}
%%%
%%%
%%%
%%%
\begin{align}
	\tDeparK{k,u}
	=
	\frac{1}{\velK{u}}
	\left(
	\BScovReq{u}
	+
	\sum_{i=1}^{\BStotal}
	\BScovK{i}
	\right),
		\quad
		k\in\BSset
		,\,\,
		u\in\userSet.
\end{align}
%%%
%%%
Let
$\userOrder{k,m}$ denote the index of the vehicle that is the $m$-th one entering \ac{RSU} $k$'s coverage area, where $m\in\userOrderIndexSet\triangleq\left\{1,\cdots,\userTot\right\}$.
Thus, $\tArrivK{k,\userOrder{k,1}}<\tArrivK{k,\userOrder{k,2}}<\cdots<\tArrivK{k,\userOrder{k,\userTot}}$.
As the $\userTot$ vehicles have different velocities and initial distances to \ac{RSU} 1, the vehicle orders for different \acp{RSU} can differ.

%\/\/\/\/\/\/\/\/\/\/\/\/\/\/\/\/\/\/\/\/\/\/\/\/\/\/\/\/\/\/\/\/\/\/\/\/\/\/\/\/\/\/\/\/\/\/\/\/\/\/\/\/\/\/\/\/\/\/\/\/\/\/\/\/\/\/\/\/\/\/\/\/\/\/\/\/\/\/\/\/\/\/\/\/\/\/\/
%***** x-axis:  [tc][bc][0.7] y-axis: [bc][tc][0.7], legend: [Bl][Bl][0.59]
%\/\/\/\/\/\/\/\/\/\/\/\/\/\/\/\/\/\/\/\/\/\/\/\/\/\/\/\/\/\/\/\/\/\/\/\/\/\/\/\/\/\/\/\/\/\/\/\/\/\/\/\/\/\/\/\/\/\/\/\/\/\/\/\/\/\/\/\/\/\/\/\/\/\/\/\/\/\/\/\/\/\/\/\/\/\/\/

\subsection{Cooperative Offloading Model}

Since each vehicle may move very fast and a single \ac{RSU} can hardly finish the computation of its entire task before the vehicle leaves its coverage area, we consider cooperative offloading among the $\BStotal$ \acp{RSU}.

\subsubsection{Computation Model}

Each vehicle's task is split into $\BStotal$ subtasks, which are offloaded to $\BStotal$ \acp{RSU}, and each \ac{RSU}.
We denote $\taskSplit{k,u}$ as the fraction of task $u$ offloaded to \ac{RSU} $k$, where
%%%
%%%
\begin{align}
	\taskSplit{k,u}
	&\geq
	0,
	\quad
	k	\in	\BSset,
	\,\,
	u	\in	\userSet,
	\label{eq:multi_data_positive}
	\\
	\sum_{k\in\BSset}
	\taskSplit{k,u}
	&=
	1,
	\quad
	u
	\in
	\userSet
	.
	\label{eq:multi_data_sum}
\end{align}
%%%
%%%
Assume that the sizes of the computation workload and the computation result of vehicle $u$'s subtask at \ac{RSU} $k$ are $\workload{u}\taskSplit{k,u}$ and $\result{u}\taskSplit{k,u}$, respectively.\footnote{
	Note that our work can be easily extended to the case where the size of the computation result is not proportional to the size of the computation workload.
}
In later sections, the task splitting will be optimized.

The CPU frequency of each \ac{RSU} can be adjusted by using dynamic voltage and frequency scaling technology \cite{WanSheWanWanLi:16}.
We denote $\BSfreqK{k,u}$ as the CPU frequency used for executing vehicle $u$'s subtask $k$ at \ac{RSU} $k$, where
%%%
%%%
\begin{align}
	\BSfreqK{k,u}
	&\geq 0
	,\quad
	k
	\in
	\BSset,
	\,\,
	u
	\in
	\userSet,
	\label{eq:multi_freq_positive}
	\\
	\BSfreqK{k,u}
	&\leq
	\BSfreqM{k}
	,\quad
	k
	\in
	\BSset,
	\,\,
	u
	\in
	\userSet.
	\label{eq:multi_freq_max_freq}
\end{align}
%%%
%%%
Here, $\BSfreqM{k}$ represents the maximum computing capability (i.e., CPU frequency) of \ac{RSU} $k$.
Then, the computation energy consumption for executing vehicle $u$'s subtask $k$ at \ac{RSU} $k$ is given by \cite{WanYan:18}:
%%%
%%%
\begin{align}
	\enerK{\CP,k,u}
	\left(\taskSplit{k,u},\BSfreqK{k,u}\right)
	&=
	\eCompCap{k}
	\workload{u}
	\taskSplit{k,u}
	\BSfreqK{k,u}^{\eCompConst{k}-1}
	,\quad
	k\in\BSset
	,\,\,
	u\in\userSet,
	\label{eq:multi_energy_cp_k_u}
\end{align}
%%%
%%%
where $\kappa_{k}> 0$ represents the effective switched capacitance depending on the chip architecture and $\CPUenergy{k}> 1$.
Moreover, the computation time for executing vehicle $u$'s subtask $k$ at \ac{RSU} $k$ is $\frac{\workload{u}\taskSplit{k,u}}{\BSfreqK{k,u}}$.

For ease of implementation, we assume that \ac{RSU} $k$ processes the $\userTot$ subtasks from the $\userTot$ vehicles one by one in the order of arrival of the $\userTot$ vehicles, i.e., $\userOrder{k,1}, \cdots, \userOrder{k,\userTot}$.
We denote $\tStartCP{k,\userOrder{k,m}}$ as the computation starting time of the $k$-th subtask of the $m$-th vehicle entering \ac{RSU} $k$'s coverage area.
Hence, we have the following computation constraints.
%%%
%%%
\begin{align}
	\tStartCP{k,\userOrder{k,m}}
	\geq
	0,
		\quad
		k\in\BSset
		,\,\,
		m\in\userOrderIndexSet
		,
%		\,\,
%		\userOrder{k,m}\in\userOrderKset{k},
	\label{eq:multi_cp_tStartCP_positive}
	\\
	\tStartCP{k,\userOrder{k,m}}
	+
	\frac
	{\workload{\userOrder{k,m}}\taskSplit{k,\userOrder{k,m}}}
	{\BSfreqK{k,\userOrder{k,m}}}
	\leq
	\tStartCP{k,\userOrder{k,m+1}},
	\nonumber
	\\
		\qquad
		k\in\BSset
		,\,\,
		m\in\userOrderIndexSet\backslash\{U\}.
%		\,\,
%		\userOrder{k,m}\in\userOrderKset{k}.
	\label{eq:multi_cp_prev_current_tStart}
\end{align}
%%%
%%%
The constraints in \eqref{eq:multi_cp_prev_current_tStart} indicate that the computation of the $m$-th vehicle's subtask $k$ at \ac{RSU} $k$ must complete before the start of the computation of the $(m+1)$-th vehicle's subtask $k$ at \ac{RSU} $k$.
We assume that the computation of vehicle $u$'s subtask at \ac{RSU} $k$ completes before vehicle $u$ enters the coverage area of \ac{RSU} $k$.
According to the assumption, we have the following computation constraint.
%%%
%%%
\begin{align}
	\tStartCP{k,\userOrder{k,m}}
	+
	\frac
	{\workload{\userOrder{k,m}}\taskSplit{k,\userOrder{k,m}}}
	{\BSfreqK{k,\userOrder{k,m}}}
	\leq
	\tArrivK{k,\userOrder{k,m}},
		\quad
		k\in\BSset
		,\,\,
		m\in\userOrderIndexSet.
	\label{eq:multi_cp_current_arriv}
\end{align}
%%%
%%%

\subsubsection{Communication Model}

After computing the $k$ subtask of vehicle $u$, \ac{RSU} $k$ transmits the computation result to vehicle $u$ when vehicle $u$ is in its coverage area.
Each \ac{RSU} has $\BSantenna$ transmit antennas, and each vehicle has a single transmit antenna.
Thus, this corresponds to \ac{MISO} transmission.
Let $\PathLoss{k}$ denote the large-scale fading power of the channel between \ac{RSU} $k$ and vehicle $u$ with the maximum link length $\linDisMaxK{k}$.
We consider a narrow band slotted system of bandwidth $\BW$ (in Hz) and adopt the block fading model for small-scale fading.
Let $\CHtra{k,u}$ denote the small-scale fading coefficients of the channel between \ac{RSU} $k$ and vehicle $u$ in an arbitrary time slot, where $\CH{k,u}\in\mathcal{C}^{\BSantenna\times 1}$.
Here, $\mathbf{H}$ denotes conjugate transpose.
Suppose that the elements of $\CH{k,u}$ are \ac{i.i.d.} according to $\mathcal{CN}\left(0,1\right)$.
For tractability, we do not adapt the transmission rate according to small scale fading (i.e., the transmission rate for each vehicle within the coverage area of each \ac{RSU} is constant) when conducting task splitting.

The transmission power limit of \ac{RSU} $k$ is denoted by $\BSpowerM{k}$.
We denote $\BSpowerK{k,u}$ as the transmission power used for transmitting vehicle $u$'s subtask $k$ from \ac{RSU} $k$ to vehicle $u$, where
%%%
%%%
\begin{align}
	\BSpowerK{k,u}
	&\geq
	0
	,\quad
	k\in\BSset
	,\,\,
	u\in\userSet,
	\label{eq:multi_power_positive}
	\\
	\BSpowerK{k,u}
	&\leq
	\BSpowerM{k}
	,\quad
	k
	\in
	\BSset
	,\,\,
	u\in\userSet.
	\label{eq:multi_power_max_power}
\end{align}
%%%
%%%
Consider an arbitrary slot. The received signal of vehicle $u$ from \ac{RSU} $k$, denoted as $\recvPower{k,u}$, is given by \cite{JinAndWeb:11}:
%%%
%%%
\begin{align}
	\recvPower{k,u}
	&=
	\CHtra{k,u}\MRTweight{k,u}
	\sqrt{\BSpowerK{k,u}\PathLoss{k}}
	\MRTsignal{k,u}
	+
	\noiseVar,
	\quad
	k\in\BSset
	,\,\,
	u\in\userSet,
	\label{eq:multi_MISO_recvPower}
\end{align}
%%%
%%%
where $\MRTsignal{k,u}$ is an information symbol with $\Expt{\left|\MRTsignal{k,u}\right|^2}=1$,
$\MRTweight{k,u}\in\mathcal{C}^{\BSantenna\times1}$ is a normalized beamforming vector with $\|\MRTweight{k,u}\|=1$, and
$\noiseVar$ is the \ac{AWGN} following $\mathcal{CN}\left(0,\noise\right)$.
From \eqref{eq:multi_MISO_recvPower}, the received \ac{SNR} of vehicle $u$ is given by:
%%%
%%%
\begin{align}
	\SNR_{k,u}
	=
	\frac
	{
		\BSpowerK{k,u}
		\left|\CH{k,u}\MRTweight{k,u}\right|^2
		\PathLoss{k}
	}
	{\noise}
	,\quad
	k\in\BSset
	,\,\,
	u\in\userSet.
	\label{eq:multi_MISO_SNR}
\end{align}
%%%
%%%
Assume that each \ac{RSU} and vehicle have perfect channel state information.
We consider \ac{MRT} beamforming, i.e.,  $\MRTweight{k,u}=\frac{\CH{k,u}}{\|\CH{k,u}\|}$, which maximizes the received signal power at vehicle $u$ \cite{JinAndWeb:11}.
The maximum achievable data rate for user $u$ at \ac{RSU} $k$ is given by:
%%%
%%%
\begin{align}
	\achievRate{\text{max},k,u}
	=
	\BW
	\log
	\left(
	1+
	\frac
	{
		\BSpowerK{k,u}
		\left\|\CH{k,u}\right\|^2
		\PathLoss{k}
	}
	{\noise}
	\right).
	\label{eq:multi_MISO_achievable_rate_max}
\end{align}

The transmission of the computation result from an \ac{RSU} to a vehicle cannot start before the vehicle enters the coverage area of the RSU and must complete before the vehicle leaves the coverage area of the \ac{RSU}.
We denote $\tStartCM{k,\userOrder{k,m}}$ and $\tCommK{k,\userOrder{k,m}}$ as the transmission starting time and the transmission time of the $k$-th subtask of the $m$-th vehicle entering \ac{RSU} $k$'s coverage area, respectively, where
%%%
%%%
\begin{align}
	\tCommK{k,\userOrder{k,m}}
	&\geq
	0,
		\quad
		k\in\BSset
		,\,\,
		m\in\userOrderIndexSet
		,
	\label{eq:multi_cm_tcm_positive}
	\\
	\tStartCM{k,\userOrder{k,m}}
	&\geq
	\tArrivK{k,\userOrder{k,m}},
		\quad
		k\in\BSset
		,\,\,
		m\in\userOrderIndexSet
		,
	\label{eq:multi_cm_tStartCM_arriv}
\\
\tStartCM{k,\userOrder{k,m}}
+
\tCommK{k,\userOrder{k,m}}
&\leq
\tDeparK{k,\userOrder{k,m}},
	\quad
	k\in\BSset
	,\,\,
	m\in\userOrderIndexSet.
\label{eq:multi_cm_current_depart}
\end{align}
%%%
%%%
Similar to the computation model, it is assumed that \ac{RSU} $k$ transmits the $\userTot$ subtasks of the $\userTot$ vehicles one by one in the order of arrival of the $\userTot$ vehicles.
Hence, we have the following communication constraints.
%%%
%%%
\begin{align}
	\tStartCM{k,\userOrder{k,m}}
	+
	\tCommK{k,\userOrder{k,m}}
	&\leq
	\tStartCM{k,\userOrder{k,m+1}},
		\,\,
		k\in\BSset
		,\,
		m\in\userOrderIndexSet\backslash\{U\}.
	\label{eq:multi_cm_prev_current_tStart}
\end{align}
%%%
%%%
The constraints in \eqref{eq:multi_cm_prev_current_tStart} indicate that \ac{RSU} $k$ starts the transmission of subtask $k$ of the $m$-th vehicle entering its coverage area after finishing the transmission of subtask $k$ of the $(m+1)$-th vehicle entering its coverage area. Besides,
\ac{RSU} $k$ has to transmit the computation result of vehicle $u$'s subtask $k$ of size $\result{u}\taskSplit{k,u}$ to vehicle $u$ within the transmission time $\tCommK{k,u}$.
Thus, the rate to transmit the computation result of size $\result{u}\taskSplit{k,u}$ to vehicle $u$ from \ac{RSU} $k$, denoted by $\achievRate{k,u}$, is given by:
%%%
%%%
\begin{align}
	\achievRate{k,u}
	=
	\frac{\result{u}\taskSplit{k,u}}{\tCommK{k,u}}
	,\quad
	k\in\BSset
	,\,\,
	u\in\userSet.
	\label{eq:multi_MISO_achievable_rate_K}
\end{align}
%%%
%%%
Besides, the communication energy consumption for transmitting the computation result from \ac{RSU} $k$ to vehicle $u$ is given by:
	%%%
	%%%
	\begin{align}
		\enerK{\CM,k,u}
		\left(\BSpowerK{k,u}, \tCommK{k,u}\right)
		&=
		\BSpowerK{k,u}
		\tCommK{k,u}
		,\quad
		k\in\BSset
		,\,\,
		u\in\userSet.
		\label{eq:multi_energy_cm_k_u}
	\end{align}
	%%%
	%%%

We impose the following successful transmission constraints.
%%%
%%%
\begin{align}
	&\Prob{
		\achievRate{\text{max},k,u}
		\geq
		\achievRate{k,u}
	}
	\geq
	\targetK{u},
		\quad
		k\in\BSset
		,\,\,
		u\in\userSet,
	\label{eq:multi_MISO_MRT_STP}
\end{align}
%%%
%%%
where $\targetK{u}$ denotes the target \ac{STP} of vehicle $u$.
Since $\|\CH{k,u}\|^2\sim\Gamma\left(\BSantenna,1\right)$, \eqref{eq:multi_MISO_MRT_STP} is equivalent to:
%%%
%%%
\begin{align}
	&
	\Prob{
		\left\|\CH{k,u}\right\|^2
		\geq
		\frac{\noise}{\BSpowerK{k,u}\PathLoss{k}}
		\left(
		2^{\frac{\result{u}\taskSplit{k,u}}{\BW\tCommK{k,u}}}
		\right)
	}
		\nonumber
		\\
		&
	=
	\MRTFunc{
		\frac{\noise}{\BSpowerK{k,u}\PathLoss{k}}
		\left(
		2^{\frac{\result{u}\taskSplit{k,u}}{\BW\tCommK{k,u}}}
		\right)
	}
	\geq
	\targetK{u},
	\label{eq:multi_MISO_MRT_STP_2}
\end{align}
%%%
%%%
where $\MRTFunc{y}\triangleq\sum_{n=0}^{\BSantenna-1} \frac{1}{n!} y^n e^{-y}$ denotes the \ac{CCDF} of $\Gamma\left(\BSantenna,1\right)$.

\subsubsection{Total Energy Consumption}
The total energy consumed at the $\BStotal$ \acp{RSU} for serving the $\userTot$ vehicles is given by:
	%%%
	%%%
	\begin{align}
			&
		\enerTot
		\left(\taskSplitSet{}, \BSfreqSet{}, \BSpowerSet{}, \tCommSet{}\right)
			\nonumber
			\\
		&=
		\sum_{k\in\BSset}
		\sum_{u\in\userSet}
		\Bigl(
		\enerK{\CP,k,u}
		\left(\taskSplit{k,u},\BSfreqK{k,u}\right)
		+
		\enerK{\CM,k,u}
		\left(\BSpowerK{k,u}, \tCommK{k,u}\right)
		\Bigr)
		\nonumber
		\\
		&=
		\sum_{k\in\BSset}
		\sum_{u\in\userSet}
		\left(
		\eCompCap{k}
		\workload{u}
		\taskSplit{k,u}
		\BSfreqK{k,u}^{\eCompConst{k}-1}
		+
		\BSpowerK{k,u}
		\tCommK{k,u}
		\right),
		\label{eq:multi_energy_objective_freq}
	\end{align}
	%%%
	%%%
	where $\taskSplitSet{}\triangleq\left(\taskSplit{k,u}\right)_{k\in\BSset,u\in\userSet}$, $\BSfreqSet{}\triangleq\left(\BSfreqK{k,u}\right)_{k\in\BSset,u\in\userSet}$, 
	$\BSpowerSet{}\triangleq\left(\BSpowerK{k,u}\right)_{k\in\BSset,u\in\userSet}$,
	$\tCommSet{}\triangleq\left(\tCommVarK{k,u}\right)_{k\in\BSset,u\in\userSet}$, 
	and the last equation is due to \eqref{eq:multi_energy_cp_k_u} and \eqref{eq:multi_energy_cm_k_u}.

We assume that all \acp{RSU} are connected to a controller, which is aware of the network parameters, $\BSantenna$, $\BStotal$, $\BSpowerM{k}$, $\BSfreqM{k}$, $\BScovK{k}$, $\linDisMaxK{k}$, $\PathLoss{k}$, $k\in\BSset$.
In the offline scenario, we further assume that the controller also knows the vehicle parameters, i.e., $\userTot$, $\velK{u}$, $\BScovReq{u}$, $\workload{u}\taskSplit{k,u}$, and $\result{u}\taskSplit{k,u}$, at time 0.
Under the above assumptions, in the offline scenario, the controller is aware of the expression of $\enerTot\left(\taskSplitSet{}, \BSfreqSet{}, \BSpowerSet{}, \tCommSet{}\right)$ at time 0.

\begin{remark}[Differences Between Proposed Cooperative Offloading Model and Cooperative Offloading Model in \cite{BozTarCor:18}]
		The merits of the proposed cooperative offloading model compared to the existing one \cite{BozTarCor:18} are summarized as follows.
		Firstly, the task splitting and computation and communication resource allocation are jointly considered in the proposed model, whereas only task splitting is considered in \cite{BozTarCor:18}.
		Secondly, the limitation on the communication capability is considered in the proposed model via \eqref{eq:multi_power_positive}, \eqref{eq:multi_power_max_power}, \eqref{eq:multi_cm_tcm_positive}\textendash\eqref{eq:multi_cm_prev_current_tStart}, but not in \cite{BozTarCor:18}.
		Thirdly, it is guaranteed to complete the execution of each subtask at each \ac{RSU} before the corresponding vehicle enters the coverage area of the \ac{RSU} in the proposed model via \eqref{eq:multi_cp_tStartCP_positive}\textendash\eqref{eq:multi_cp_current_arriv}, but not in \cite{BozTarCor:18} (where outage of task is allowed).
		Finally, the computation and communication operation orders for the subtasks of all moving vehicles are reflected in the newly specified constraints \eqref{eq:multi_cp_prev_current_tStart}, \eqref{eq:multi_cp_current_arriv}, \eqref{eq:multi_cm_current_depart}, and \eqref{eq:multi_cm_prev_current_tStart}, but not in \cite{BozTarCor:18}.
		To the best of our knowledge, this is the first cooperative offloading model for edge computing-enabled vehicular networks which reflects the computation and communication resource limitation, allows efficient utilization of computation and communication resources, and provides performance guarantees for latency-sensitive and computation-intensive applications.	 
\end{remark}

\section{Total Energy Minimization In Offline Scenario}
\label{sec:problem_formulation}
In this section, we consider the total energy minimization in the offline scenario.
First, we formulate the total energy minimization as a non-convex problem. Then, we obtain a globally optimal solution for the multi-vehicle case. Finally, we obtain a globally optimal solution for the single-vehicle case.

\subsection{Problem Formulation}
In the offline scenario, we would like to minimize the total energy consumption $\enerTot\left(\taskSplitSet{}, \BSfreqSet{}, \BSpowerSet{}, \tCommSet{}\right)$ in \eqref{eq:multi_energy_objective_freq} by optimizing 
the task splitting factor $\taskSplitSet{}$, the CPU frequency allocation $\BSfreqSet{}$, the transmission power allocation $\BSpowerSet{}$, the transmission time allocation $\tCommSet{}$, the computation starting time allocation $\tStartCPSet{}\triangleq\left(\tStartCP{k,u}\right)_{k\in\BSset,u\in\userSet}$, and the transmission starting time allocation $\tStartCMSet\triangleq\left(\tStartCM{k,u}\right)_{k\in\BSset,u\in\userSet}$.
The optimization problem is formulated as follows.
%%%%%%%%%%%%%%%%%%%%%%%%%%%%%%%%%%%%%%%%%%%%%%%%%%%%%%%%%%%%%%%%%%%%%%%%%%%%%
%----------------------------------------------------------------------------
\begin{problem}[Total Energy Minimization in Offline Scenario]
	\begin{alignat}{2}
		\enerOpt
		\triangleq
		\min_{\taskSplitSet{}, \BSfreqSet{}, \BSpowerSet{}, \tCommSet{}, \tStartCPSet{}, \tStartCMSet}\quad
		&  
		\enerTot
		\left(\taskSplitSet{}, \BSfreqSet{}, \BSpowerSet{}, \tCommSet{}\right)
		\nonumber
		%\label{eq:ObjFunc_single_original}
		\\
		\text{s.t.}\quad\,\,\,
		& 
		% (3)
		\eqref{eq:multi_data_positive},
		% (4)
		\eqref{eq:multi_data_sum},
		% (5)
		\eqref{eq:multi_freq_positive},
		% (6)
		\eqref{eq:multi_freq_max_freq},
		% (8)
		\eqref{eq:multi_cp_tStartCP_positive},
		% (9)
		\eqref{eq:multi_cp_prev_current_tStart},
		% (10)
		\eqref{eq:multi_cp_current_arriv},
		% (11)
		\eqref{eq:multi_power_positive},
		\nonumber
		\\
		&
		% (12)
		\eqref{eq:multi_power_max_power},
		% (16)
		\eqref{eq:multi_cm_tcm_positive},
		% (17)
		\eqref{eq:multi_cm_tStartCM_arriv},
		% (18)
		\eqref{eq:multi_cm_current_depart},	
		% (19)
		\eqref{eq:multi_cm_prev_current_tStart},
		% (23)
		\eqref{eq:multi_MISO_MRT_STP_2},
		\nonumber
	\end{alignat}
	\label{prob:multi_energy_min_freq}
\end{problem}
%----------------------------------------------------------------------------
%%%%%%%%%%%%%%%%%%%%%%%%%%%%%%%%%%%%%%%%%%%%%%%%%%%%%%%%%%%%%%%%%%%%%%%%%%%%%
where $\enerTot\left(\taskSplitSet{}, \BSfreqSet{}, \BSpowerSet{}, \tCommSet{}\right)$ is given by \eqref{eq:multi_energy_objective_freq}.
Let $\left(\taskSplitOptSet, \BSfreqOptSet, \BSpowerOptSet, \tCommOptSet, \tStartCPOptSet, \tStartCMOptSet\right)$ denote an optimal solution of Problem \ref{prob:multi_energy_min_freq}.

Given the network parameters and the vehicle parameters, the controller can solve Problem \ref{prob:multi_energy_min_freq}.
In Problem \ref{prob:multi_energy_min_freq}, the objective function is non-convex,
the inequality constraint functions in
\eqref{eq:multi_cp_prev_current_tStart}, %(9)
\eqref{eq:multi_cp_current_arriv}, %(10)
and \eqref{eq:multi_MISO_MRT_STP_2} %(25)
are non-convex,
the inequality constraint functions in 
\eqref{eq:multi_data_positive}, %(3)
\eqref{eq:multi_freq_positive}, %(5)
\eqref{eq:multi_freq_max_freq}, %(6)
\eqref{eq:multi_cp_tStartCP_positive}, %(8)
\eqref{eq:multi_power_positive}, %(12)
\eqref{eq:multi_power_max_power}, %(13)
\eqref{eq:multi_cm_tcm_positive}, %(14)
\eqref{eq:multi_cm_tStartCM_arriv}, %(15)
\eqref{eq:multi_cm_prev_current_tStart}, %(16)
and \eqref{eq:multi_cm_current_depart} %(17)
are convex,
and the equality constraint function in 
\eqref{eq:multi_data_sum} %(4)
is affine.
Therefore, Problem \ref{prob:multi_energy_min_freq} is a non-convex problem with $6\BStotal\userTot$ variables and $\left(13\BStotal\userTot-2\BStotal+\userTot\right)$ constraints.
In general, it is hard to obtain a globally optimal solution for a non-convex problem analytically or numerically with effective and efficient methods.
Besides, Problem~\ref{prob:multi_energy_min_freq} is more challenging than those for cooperative offloading in the static scenario \cite{GuoSonCuiLiuJi:17,LiaLiuLokHua:19,WanYaoZheSunSon:18,WanXuWanCui:18,BiYuaDuaZhoAbu:21,LiXuZhoLuHan:20,CheTenSunLiuWan:18,WanSheWanWanLi:16,DinTanLaQue:17,GuWanLiuFanFanZho:18} and the vehicular networks \cite{BozTarCor:18}.
This is because Problem~\ref{prob:multi_energy_min_freq} involves considerably more variables and constraints to specify the limitation on the computation and communication capability, and the computation and communication operation orders for the subtasks of all moving vehicles.
In Section \ref{sec:optimal_solution_off-multi} and Section \ref{sec:optimal_solution_off-single}, we solve Problem~\ref{prob:multi_energy_min_freq} in the multi-vehicle case $\left(\userTot>1\right)$ and single-vehicle case $\left(\userTot=1\right)$, respectively.
Specifically, in each case, we transform Problem \ref{prob:multi_energy_min_freq} into a convex problem with fewer variables and constraints by characterizing and utilizing an optimality property and obtain a globally optimal solution using convex optimization techniques.

\subsection{Optimal Solution in Multi-Vehicle Case}
\label{sec:optimal_solution_off-multi}

In this subsection, we consider the multi-vehicle case, i.e., $\userTot>1$.
First, we characterize an optimality property of Problem \ref{prob:multi_energy_min_freq}.
%%%%%%%%%%%%%%%%%%%%%%%%%%%%%%%%%%%%%%%%%%%%%%%%%%%%%%%%%%%%%%%%%%%%%%%%%%%%%
%----------------------------------------------------------------------------
\begin{lemma}[Optimal Transmission Power Allocation]
	An optimal solution of Problem \ref{prob:multi_energy_min_freq} satisfies:
	%%%
	%%%
	\begin{align}
		\BSpowerOptK{k,u}
		=
		\frac{\noise}{\PathLoss{k}\MRTFuncInv{\targetK{u}}}
		\left(
		2^{\frac{\result{u}\taskSplitOptK{k,u}}{\BW\tCommOptK{k,u}}}
		-
		1
		\right)
				,\quad
				k\in\BSset
				,\,\,
				u\in\userSet,
	\end{align}
	%%%
	%%%
	where $\MRTFuncInv{\cdot}$ denotes the inverse function of $\MRTFunc{\cdot}$.\footnote{$\MRTFuncInv{\cdot}$ can be numerically computed \cite{BarSte:12}.}
	\label{lemma:multi_opt_power}
\end{lemma}
%----------------------------------------------------------------------------
\begin{IEEEproof}
	We equivalently convert the constraints in \eqref{eq:multi_MISO_MRT_STP_2} to:
	%%%
	%%%
	\begin{align}
		\BSpowerK{k,u}
		\geq
		\frac{\noise}{\PathLoss{k}\MRTFuncInv{\targetK{u}}}
		\left(
		2^{\frac{\result{u}\taskSplit{k,u}}{\BW\tCommK{k,u}}}
		-
		1
		\right),
			\quad
			k\in\BSset
			,\,\,
		u\in\userSet.
		\label{eq:multi_MRT_rate_constraint}
	\end{align}
	%%%
	%%%
	Besides, for given $\taskSplit{k,u}$, $\BSfreqK{k,u}$, $\tCommK{k,u}$, $\tStartCP{k,u}$, and $\tStartCM{k,u}$, $\enerTot\left(\taskSplitSet{}, \BSfreqSet{}, \BSpowerSet{}, \tCommSet{}\right)$ in \eqref{eq:multi_energy_objective_freq} increases with $\BSpowerK{k,u}$, $k\in\BSset, u\in\userSet$.
	Thus, by contradiction, we can show that the inequality constraints in \eqref{eq:multi_MRT_rate_constraint} are active at $\left(\taskSplitOptSet, \BSfreqOptSet, \BSpowerOptSet, \tCommOptSet, \tStartCPOptSet, \tStartCMOptSet\right)$.
	Therefore, we complete the proof of Lemma \ref{lemma:multi_opt_power}.
\end{IEEEproof}
%----------------------------------------------------------------------------
%%%%%%%%%%%%%%%%%%%%%%%%%%%%%%%%%%%%%%%%%%%%%%%%%%%%%%%%%%%%%%%%%%%%%%%%%%%%%

Next, we equivalently convert the non-convex problem in Problem \ref{prob:multi_energy_min_freq} to a convex problem with fewer variables and constraints by changing variables and using Lemma~\ref{lemma:multi_opt_power}.
We introduce the computation time for executing vehicle $u$'s subtask $k$ with the workload of size $\CPUunitK{u}\taskSplit{k,u}$ at \ac{RSU} $k$, denoted by $\tCompK{k,u}$, which satisfies:
%%%
%%%
\begin{align}
	\tCompK{k,u}
	=
	\frac
	{\workload{u}\taskSplit{k,u}}
	{\BSfreqK{k,u}},
	\quad
	k\in\BSset
	,\,\,
	u\in\userSet
	.
	\label{eq:multi_tcp_def}
\end{align}
%%%
%%%
Let $\tCompSet{}\triangleq\left(\tCompK{k,u}\right)_{k\in\BSset,u\in\userSet}$.
Define
\begin{align}
	\enerTotTild
	\left(\taskSplitSet{}, \tCompSet, \tCommSet{}\right)
	\triangleq&
	\sum_{k=1}^{\BStotal}
	\sum_{u=1}^{\userTot}
	\left(
	\enerKTild{\CP,k,u}
	\left(\taskSplit{k,u},\tCompK{k,u}\right)
			\right.
			\nonumber
			\\
			&
			\left.
	+
	\enerKTild{\CM,k,u}
	\left(\taskSplit{k,u}, \tCommK{k,u}\right)
	\right),
	\label{eq:multi_energy_objective_x_tcp_tcm}
\end{align}
%%%
%%%
where
%%%
\begin{align}
	\enerKTild{\CP,k,u}
	\left(\taskSplit{k,u},\tCompK{k,u}\right)
	&\triangleq
	\eCompCap{k}
	\workload{u}^{\eCompConst{k}}
	\taskSplit{k,u}^{\eCompConst{k}}
	\tCompK{k,u}^{1-\eCompConst{k}},
	\label{eq:multi_energy_cp_x_tcp}
	\\
	\enerKTild{\CM,k,u}
	\left(\taskSplit{k,u}, \tCommK{k,u}\right)
	&\triangleq
	\frac
	{\noise\tCommK{k,u}}
	{\PathLoss{k}\MRTFuncInv{\targetK{u}}}
	\left(
	2^{\frac{\result{u}\taskSplit{k,u}}{\BW\tCommK{k,u}}}
	-
	1
	\right).
	\label{eq:multi_energy_cm_x_tcm}
\end{align}
%%%
%%%
By \eqref{eq:multi_tcp_def} and Lemma \ref{lemma:multi_opt_power}, we can transform Problem \ref{prob:multi_energy_min_freq} into the following problem with $5\BStotal\userTot$ variables and $\left(9\BStotal\userTot-\BStotal+\userTot\right)$ constraints.
%%%%%%%%%%%%%%%%%%%%%%%%%%%%%%%%%%%%%%%%%%%%%%%%%%%%%%%%%%%%%%%%%%%%%%%%%%%%%
%----------------------------------------------------------------------------
\begin{problem}[Equivalent Problem of Problem \ref{prob:multi_energy_min_freq} in Multi-Vehicle Case]
	\begin{alignat}{2}
		\min_{
			\substack{
				\taskSplitSet{}, \tCompSet, \tCommSet{},\\
				\tStartCPSet{}, \tStartCMSet
			}
		}
		\,\,
		&  
		\enerTotTild
		\left(\taskSplitSet{}, \tCompSet, \tCommSet{}\right)
		\nonumber
		%\label{eq:ObjFunc_single_original}
		\\
		\text{s.t.}\quad
		& 
		% (3)
		\eqref{eq:multi_data_positive},
		% (4)
		\eqref{eq:multi_data_sum},
		% (8)
		\eqref{eq:multi_cp_tStartCP_positive},
		% (15)
		\eqref{eq:multi_cm_tStartCM_arriv},
		% (16)
		\eqref{eq:multi_cm_current_depart},
		% (17)
		\eqref{eq:multi_cm_prev_current_tStart},
		\nonumber
		\\
		&
		\tCommK{k,u}
		%		\hspace{-0.5mm}
		-
		%		\hspace{-0.5mm}
		\frac
		{
			\result{u}\taskSplit{k,u}
		}
		{
			\BW
			\log_2
			\left(
			1
			+
			\frac{\BSpowerM{k}\PathLoss{k}\MRTFuncInv{\targetK{u}}}{\noise}
			\right)
		}
		\geq
		0,
				\nonumber
				\\
		&\qquad\qquad\qquad\qquad\qquad
		\quad
		k\in\BSset
		,\,\,
		u\in\userSet
		,
		\label{eq:multi_MRT_optimal_power}
		\\
		&
		\taskSplit{k,u}
		-
		\frac{\BSfreqM{k}}{\workload{u}}
		\tCompK{k,u}
		\leq
		0,
		\qquad
		k\in\BSset
		,\,\,
		u\in\userSet
		,
		\label{eq:multi_cp_tcp_max_freq}
		\\
		&
		\tStartCP{k,\userOrder{k,m-1}}
		+
		\tCompK{k,\userOrder{k,m-1}}
		\leq
		\tStartCP{k,\userOrder{k,m}},
					\nonumber\\
					&
		\qquad
		k\in\BSset
		,\,\,
		m\in\userOrderIndexSet
		,
		\label{eq:multi_cp_tcp_prev_current_tStart}
		\\
		&
		\tStartCP{k,\userOrder{k,m}}
		+
		\tCompK{k,\userOrder{k,m}}
		\leq
		\tArrivK{k,\userOrder{k,m}},
		\quad
		k\in\BSset
		,\,\,
		m\in\userOrderIndexSet
		,
		\label{eq:multi_cp_tcp_current_arriv}
	\end{alignat}
	\label{prob:multi_energy_min_tcp}
\end{problem}
%----------------------------------------------------------------------------
%%%%%%%%%%%%%%%%%%%%%%%%%%%%%%%%%%%%%%%%%%%%%%%%%%%%%%%%%%%%%%%%%%%%%%%%%%%%%
where $\enerTotTild\left(\taskSplitSet{}, \tCompSet, \tCommSet{}\right)$ is given by \eqref{eq:multi_energy_objective_x_tcp_tcm}.
Let $\left(\taskSplitOptTmpSet, \tCompOptSet, \tCommOptTmpSet, \tStartCPOptTmpSet, \tStartCMOptTmpSet\right)$ denote an optimal solution of Problem \ref{prob:multi_energy_min_tcp}.

Then, we show that Problem \ref{prob:multi_energy_min_freq} is equivalent to Problem \ref{prob:multi_energy_min_tcp} with fewer variables and constraints.
%%%%%%%%%%%%%%%%%%%%%%%%%%%%%%%%%%%%%%%%%%%%%%%%%%%%%%%%%%%%%%%%%%%%%%%%%%%%%
%----------------------------------------------------------------------------
\begin{theorem}[Equivalence between Problem \ref{prob:multi_energy_min_freq} and Problem \ref{prob:multi_energy_min_tcp} in Multi-Vehicle Case]
	In the multi-vehicle case, the solutions of Problem \ref{prob:multi_energy_min_freq} and Problem \ref{prob:multi_energy_min_tcp} satisfy:
	%%%
	%%%
	\begin{gather}
		\taskSplitOptTmpSet=\taskSplitOptSet, \quad
		\tCompOptSet=\left(\frac{\CPUunitK{u}\taskSplitOptK{k,u}}{\BSfreqOptK{k,u}}\right)_{k\in\BSset,u\in\userSet}, \quad
		\tCommOptTmpSet=\tCommOptSet, \quad
				\nonumber
				\\
		\tStartCPOptTmpSet=\tStartCPOptSet,\quad
		\tStartCMOptTmpSet=\tStartCMOptSet.
		\nonumber
	\end{gather}
	%%%
	%%%
	\label{theorem:off-multi_convex_transform}
\end{theorem}
%----------------------------------------------------------------------------
\begin{IEEEproof}
	See Appendix~\ref{app:theo_multi_convex_transform}.
\end{IEEEproof}
%----------------------------------------------------------------------------
%%%%%%%%%%%%%%%%%%%%%%%%%%%%%%%%%%%%%%%%%%%%%%%%%%%%%%%%%%%%%%%%%%%%%%%%%%%%%

All constraints in Problem \ref{prob:multi_energy_min_tcp} are convex.
In the following lemma, we show that the objective function of Problem \ref{prob:multi_energy_min_tcp} is also convex.
\begin{lemma}[Convexity of Objective Function of Problem~\ref{prob:multi_energy_min_tcp}]
	$\enerTotTild\left(\taskSplitSet{}, \tCompSet, \tCommSet{}\right)$ in \eqref{eq:multi_energy_objective_x_tcp_tcm} is convex in $\left(\taskSplitSet{}, \tCompSet, \tCommSet{}\right)$.
	\label{lemma:conv_obj_multi}
\end{lemma}
%----------------------------------------------------------------------------
\begin{IEEEproof}
	Define $\perspecCP{k,u}(y)=\eCompCap{k}\workload{u}^{\eCompConst{k}}y^{\eCompConst{k}}$, which is a convex function of $y$.
	Since the perspective operation preserves convexity, $\enerKTild{\CP,k,u}\left(\taskSplit{k,u},\tCompK{k,u}\right)=\tCompK{k,u}\perspecCP{k,u}\left(\taskSplit{k,u}/\tCompK{k,u}\right)$ is also a convex function of $\left(\taskSplit{k,u},\tCompK{k,u}\right)$.
	Similarly, we can show that $\enerKTild{\CM,k,u}\left(\taskSplit{k,u}, \tCommK{k,u}\right)$ is a convex function of $\left(\taskSplit{k,u},\tCompK{k,u}\right)$.
	Therefore, $\enerTotTild\left(\taskSplitSet{}, \tCompSet, \tCommSet{}\right)$ is a convex function of $\left(\taskSplit{k,u},\tCompK{k,u}\right)$.
\end{IEEEproof}
%----------------------------------------------------------------------------

By Lemma~\ref{lemma:conv_obj_multi}, we know that Problem~\ref{prob:multi_energy_min_tcp} is a convex problem, and hence can be solved optimally using an interior-point method of computational complexity $\mathcal{O}\left(\BStotal^3 \userTot^3\right)$.

\subsection{Optimal Solution in Single-Vehicle Case}
\label{sec:optimal_solution_off-single}
In this subsection, we consider the single-vehicle case, i.e., $\userTot=1$.
Note that in the single-vehicle case, we remove index $u$ for simplicity.
First, consider the following problem with $4\BStotal$ variables and $\left(9\BStotal+1\right)$ constraints.
%%%%%%%%%%%%%%%%%%%%%%%%%%%%%%%%%%%%%%%%%%%%%%%%%%%%%%%%%%%%%%%%%%%%%%%%%%%%%
%----------------------------------------------------------------------------
\begin{problem}[Equivalent Problem of Problem~\ref{prob:multi_energy_min_freq} in Single-Vehicle Case]
	\begin{alignat}{2}
		\min_{\taskSplitSet{}, \BSfreqSet{}, \BSpowerSet{}, \tCommSet{}}\quad
		&  
		\enerTot
		\left(\taskSplitSet{}, \BSfreqSet{}, \BSpowerSet{}, \tCommSet{}\right)
		\nonumber
		%\label{eq:ObjFunc_single_original}
		\\
		\text{s.t.}\quad\,\,\,
		& 
		% (3)
		\eqref{eq:multi_data_positive},
		% (4)
		\eqref{eq:multi_data_sum},
		% (5)
		\eqref{eq:multi_freq_positive},
		% (6)
		\eqref{eq:multi_freq_max_freq},
		% (11)
		\eqref{eq:multi_power_positive},
		% (12)
		\eqref{eq:multi_power_max_power},
		% (16)
		\eqref{eq:multi_cm_tcm_positive},
		% (23)
		\eqref{eq:multi_MISO_MRT_STP_2},
		\nonumber
		\\
		&
		\workload{}\taskSplit{k}
		-
		\BSfreqK{k}
		\tArrivK{k}
		\leq
		0
		,\quad
		k\in\BSset
		\label{eq:single_cp_arriv}
		\\
		&
		\tCommK{k}
		\leq
		\tDeparK{k}
		-
		\tArrivK{k}
		,\quad
		k\in\BSset
		\label{eq:single_cm_dep_arriv}.
	\end{alignat}
	\label{prob:single_energy_min}
\end{problem}
%----------------------------------------------------------------------------
%%%%%%%%%%%%%%%%%%%%%%%%%%%%%%%%%%%%%%%%%%%%%%%%%%%%%%%%%%%%%%%%%%%%%%%%%%%%%
\noindent Let $\left(\taskSplitOptSingle, \BSfreqOptSingle, \BSpowerOptSingle, \tCommOptSingle\right)$ denote an optimal solution of Problem \ref{prob:single_energy_min}.

We can show that Problem \ref{prob:multi_energy_min_freq} is equivalent to Problem \ref{prob:single_energy_min} with fewer variables and constraints.
%%%%%%%%%%%%%%%%%%%%%%%%%%%%%%%%%%%%%%%%%%%%%%%%%%%%%%%%%%%%%%%%%%%%%%%%%%%%%
%----------------------------------------------------------------------------
\begin{theorem}[Equivalence between Problem \ref{prob:multi_energy_min_freq} and Problem \ref{prob:single_energy_min} in Single-Vehicle Case]
	In the single-vehicle case, $\tStartCPOptK{k}=0$ and $\tStartCMOptK{k}=\tArrivK{k}$. Furthermore, the optimal solutions of Problem \ref{prob:multi_energy_min_freq} and Problem \ref{prob:single_energy_min} satisfy:
	%%%
	%%%
	\begin{gather}
		\taskSplitOptSingle=\taskSplitOptSet, \quad
		\BSfreqOptSingle=\BSfreqOptSet, \quad
		\BSpowerOptSingle=\BSpowerOptSet, \quad
		\tCommOptSingle=\tCommOptSet.
		\label{eq:single_equiv_sol}
	\end{gather}
	%%%
	%%%
	\label{theorem:single_equivalence}
\end{theorem}
%----------------------------------------------------------------------------
\begin{IEEEproof}
		See Appendix~\ref{app:theo_single_equiv}.
\end{IEEEproof}
%----------------------------------------------------------------------------
%%%%%%%%%%%%%%%%%%%%%%%%%%%%%%%%%%%%%%%%%%%%%%%%%%%%%%%%%%%%%%%%%%%%%%%%%%%%%

Based on Theorem~\ref{theorem:single_equivalence}, we can solve Problem~\ref{prob:single_energy_min} instead of Problem~\ref{prob:multi_energy_min_freq}.
Adopting the primal decomposition method, we decompose Problem~\ref{prob:single_energy_min}
into several subproblems, each with $1$ variable and $2$ constraints, controlled by a master problem with $\BStotal$ variables and $\left(3\BStotal+1\right)$ constraints \cite{PalChi:06}.
%%%%%%%%%%%%%%%%%%%%%%%%%%%%%%%%%%%%%%%%%%%%%%%%%%%%%%%%%%%%%%%%%%%%%%%%%%%%%
%----------------------------------------------------------------------------
\begin{problem}[Master Problem - Task splitting]
	%%%
	%%%
	\begin{alignat}{2}
		\min_{\taskSplitSet{}} \quad 
		& 
		\sum_{k=1}^{\BStotal}
		\left(
%		\enerKTild{\CP,k}
%		\left(\taskSplit{k}\right)
		\eCompCap{k}
		\workload{}^{\eCompConst{k}}
		\left(\tArrivK{k}\right)^{1-\eCompConst{k}}
		\taskSplit{k}^{\eCompConst{k}}
		+
		\enerK{\CM,k}^{*}
		\left(\taskSplit{k}\right)
		\right)
		\nonumber
		\\
		\text{s.t.} \quad
		&
		% (3)
		\eqref{eq:multi_data_positive},
		% (4)
		\eqref{eq:multi_data_sum},
		\nonumber
		\\	
		&
		\taskSplit{k}
		%		\hspace{-0.5mm}
		\leq
		%		\hspace{-0.5mm}
		\frac{\BSfreqM{k}\tArrivK{k}}{\workload{}}
		,\,\,
		k
		\in
		\BSset,
		\label{eq:single_opt_freq_max_freq}
		\\
		&
		\taskSplit{k}
%		\hspace{-0.5mm}
		\leq
%		\hspace{-0.5mm}
		\frac{\BW\left(\tDeparK{k}
%			\hspace{-0.8mm}
			-
%			\hspace{-0.5mm}
			\tArrivK{k}\right)}{\result{}}
		\log_2
		\left(
%		\hspace{-0.5mm}
		1+
		\frac
		{\BSpowerM{k}\PathLoss{k}\MRTFuncInv{\targetK{}}}
		{\noise}
%		\hspace{-0.5mm}
		\right),
		\nonumber
		\\
		&\qquad\qquad\qquad\qquad\qquad\qquad
		\quad
		k\in\BSset.
		\label{eq:single_opt_tcm_max_power_feasibility}
	\end{alignat}
	%%%
	%%%
	Let $\taskSplitOptDDagSet{}$ denote an optimal solution of Problem \ref{prob:single_master_data}.
	\label{prob:single_master_data}
\end{problem}
%----------------------------------------------------------------------------
%%%%%%%%%%%%%%%%%%%%%%%%%%%%%%%%%%%%%%%%%%%%%%%%%%%%%%%%%%%%%%%%%%%%%%%%%%%%%

%%%%%%%%%%%%%%%%%%%%%%%%%%%%%%%%%%%%%%%%%%%%%%%%%%%%%%%%%%%%%%%%%%%%%%%%%%%%%
%----------------------------------------------------------------------------
\begin{problem}[Subproblem - Power Allocation at \ac{RSU} $k\in\BSset$]
	For any $\taskSplit{k}$,
	%
	%
	%\begin{subequations}
	\begin{alignat}{2}
		\enerK{\CM,k}^{*}\left(\taskSplit{k}\right)
		\triangleq
		\min_{\BSpowerK{k}} \quad
		& 
%		\enerKTild{\CM,k}
%		\left(\taskSplit{k},\BSpowerK{k}\right)
		\frac
		{\result{}\taskSplit{k}\BSpowerK{k}}
		{\BW\log_2 \left(1+\frac{\BSpowerK{k}\PathLoss{k}}{\noise}\MRTFuncInv{\targetK{}}\right)}
		\label{eq:single_sub_obj}
		\\
		\text{s.t.} \quad
		&
		% (13)
		\eqref{eq:multi_power_max_power},
		\nonumber
		\\
		&
		\BSpowerK{k}
		\geq
		\frac{\noise}{\PathLoss{k}\MRTFuncInv{\targetK{}}}
		\left(
		2^{\frac{\result{}\taskSplit{k}}{\BW\left(\tDeparK{k}-\tArrivK{k}\right)}}
		-
		1
		\right).
		%		,\quad k\in\BSset
		\label{eq:single_opt_tcm_rate_constraint}
	\end{alignat}
	%\end{subequations}
	%
	%
	Let $\BSpowerOptDDagK{k}\left(\taskSplit{k}\right)$ denote an optimal solution of Problem \ref{prob:single_sub_power}.
	\label{prob:single_sub_power}
\end{problem}
%----------------------------------------------------------------------------

We can show that Problem \ref{prob:single_energy_min} is equivalent to Problem \ref{prob:single_master_data} and Problem \ref{prob:single_sub_power} which have fewer variables and constraints.
%%%%%%%%%%%%%%%%%%%%%%%%%%%%%%%%%%%%%%%%%%%%%%%%%%%%%%%%%%%%%%%%%%%%%%%%%%%%%
%----------------------------------------------------------------------------
\begin{theorem}[Equivalence between Problem \ref{prob:single_energy_min} and Problems \ref{prob:single_master_data} and \ref{prob:single_sub_power}]
	\label{theorem:single_p456_equiv}
	The solution of Problem \ref{prob:single_energy_min} and the solution of Problem \ref{prob:single_master_data} and Problem \ref{prob:single_sub_power} satisfy:
	%%%
	%%%
	\begin{gather}
		\taskSplitOptSingle=\taskSplitOptDDagSet, \quad
		\BSpowerOptSingle
		=
		\left(\BSpowerOptDDagK{k}\left(\taskSplitOptDDagK{k}\right)\right)_{k\in\BSset}, \quad
%		\BSpowerOptDDagSet\left(\taskSplitOptDDagSet\right)
%		\nonumber
%		\\
%		f^{\dagger}_k=
		\BSfreqOptSingle
		=
		\left(\frac
		{
			\workload{}\taskSplitOptDDagK{k}
		}
		{
			\tArrivK{k}
		}\right)_{k\in\BSset}
		,\quad
		%%%%%%%%%%%%%%%
		\nonumber
		\\
		\tCommOptSingle
		=
		\left(
		\frac
		{\result{}\taskSplitOptDDagK{k}}
		{\BW\log_2 \left(1+\frac{\PathLoss{k}\BSpowerOptDDagK{k}\left(\taskSplitOptDDagK{k}\right)}{\noise}\MRTFuncInv{\targetK{}}\right)}
		\right)_{k\in\BSset}.
		\nonumber
	\end{gather}
	%%%
	%%%
\end{theorem}
%----------------------------------------------------------------------------
\begin{IEEEproof}
	See Appendix~\ref{app:theo2}.
\end{IEEEproof}
%----------------------------------------------------------------------------
%%%%%%%%%%%%%%%%%%%%%%%%%%%%%%%%%%%%%%%%%%%%%%%%%%%%%%%%%%%%%%%%%%%%%%%%%%%%%

Due to the equivalence shown in Theorem \ref{theorem:single_p456_equiv}, we also use $\taskSplitOptSet$ and $\left(\BSpowerKopt{k}
\left(\taskSplit{k}\right)\right)_{k\in\BSset}$ to represent the optimal solutions of Problem \ref{prob:single_master_data} and Problem \ref{prob:single_sub_power}, respectively, with a slight abuse of notation.
Furthermore, based on Theorem~\ref{theorem:single_p456_equiv}, we can solve Problem~\ref{prob:single_master_data} and Problem~\ref{prob:single_sub_power} instead of Problem~\ref{prob:single_energy_min}.

First, we obtain a closed-form expression of an optimal solution for each subproblem in Problem \ref{prob:single_sub_power}.
%%%%%%%%%%%%%%%%%%%%%%%%%%%%%%%%%%%%%%%%%%%%%%%%%%%%%%%%%%%%%%%%%%%%%%%%%%%%%
%----------------------------------------------------------------------------
\begin{lemma}[Optimal Solution of Problem \ref{prob:single_sub_power}]
	%	The optimal solution $\BSpowerKoptSet\left(\taskSplitSet{}\right)$ is given by
	\begin{align}
		\BSpowerKopt{k}
		\left(\taskSplit{k}\right)
		=
		\frac{\noise}{\PathLoss{k}\MRTFuncInv{\targetK{}}}
		\left(
		2^{\frac{\result{}\taskSplit{k}}{\BW\left(\tDeparK{k}-\tArrivK{k}\right)}}
		-
		1
		\right),\quad k\in\BSset.
		\label{eq:single_opt_power}
	\end{align}
	\label{lemma:opt_sol_p4}
\end{lemma}
%----------------------------------------------------------------------------
\begin{IEEEproof}
	For given $\taskSplit{k}$, the objective function decreases with $\BSpowerK{k}$.
	Thus, by contradiction, we can show that the inequality constraint in \eqref{eq:single_opt_tcm_rate_constraint} is active at $\BSpowerKopt{k}$.
	Therefore, we complete the proof of Lemma \ref{lemma:opt_sol_p4}.	
\end{IEEEproof}
%----------------------------------------------------------------------------
%%%%%%%%%%%%%%%%%%%%%%%%%%%%%%%%%%%%%%%%%%%%%%%%%%%%%%%%%%%%%%%%%%%%%%%%%%%%%

Next, we solve Problem \ref{prob:single_master_data}.
By \eqref{eq:single_sub_obj} and Lemma~\ref{lemma:opt_sol_p4}, we have:
%From Lemma \ref{lemma:opt_sol_p4}, we can calculate $\enerK{\CM,k}^{*}\left(\taskSplit{k}\right)=\enerKTild{\CM,k}
%	\left(\taskSplit{k},\BSpowerKopt{k}\left(\taskSplit{k}\right)\right)$ as follows.
\begin{align}
	\enerK{\CM,k}^{*}\left(\taskSplit{k}\right)
%	=\enerKTild{\CM,k}
%	\left(\taskSplit{k},\BSpowerKopt{k}\left(\taskSplit{k}\right)\right)
	%	\nonumber
	%	\\
	%	&
	=
	\frac
	{\noise\left(\tDeparK{k}-\tArrivK{k}\right)}
	{\PathLoss{k}\MRTFuncInv{\targetK{}}}
	\left(
	2^{\frac{\compResultK{k}}{\BW\left(\tDeparK{k}-\tArrivK{k}\right)}}
	-
	1
	\right).
\end{align}
Since the objective function is convex and all constraints are affine functions, Problem \ref{prob:single_master_data} is convex.
As strong duality holds for Problem~\ref{prob:single_master_data}, we can obtain a semi-closed-form optimal solution of Problem \ref{prob:single_master_data} using \ac{KKT} conditions \cite{BoyVan:B04}, which is summarized as follows.

%%%%%%%%%%%%%%%%%%%%%%%%%%%%%%%%%%%%%%%%%%%%%%%%%%%%%%%%%%%%%%%%%%%%%%%%%%%%%
%----------------------------------------------------------------------------
\begin{lemma}[Optimal Solution of Problem \ref{prob:single_master_data}]
	\label{lemma:optimal_x}
	%%%
	%%%
\begin{align}
	\taskSplitOptK{k}
	=&
	\min
	\hspace{-0.5mm}
	\left[
	\frac{\BW\left(\tDeparK{k}
		\hspace{-0.8mm}
		-
		\hspace{-0.5mm}
		\tArrivK{k}\right)}{\result{}}
	\log_2
	\left(
	1+
	\frac
	{\BSpowerM{k}\PathLoss{k}\MRTFuncInv{\targetK{}}}
	{\noise}
	\right)
	,
	\right.
	\nonumber
	\\
	&\qquad\,\,
	\left.
	\,\,
	\frac{\BSfreqM{k}\tArrivK{k}}{\workload{}}
	,
	\,\,
	\max
	\hspace{-0.5mm}
	\left\{
	\KKTFuncInv{k}{\KKTeqOpt{}},
	\,\,
	0
	\right\}
	\right],
	\,\,
	k\in\BSset,
	\label{eq:single_opt_data}
\end{align}
	%%%
	%%%
	where $\KKTeqOpt{}$ satisfies $\sum_{k=1}^{\BStotal}\taskSplitOptK{k}=1$ with $\taskSplitOptK{k}$ given by \eqref{eq:single_opt_data} and 
	$\KKTFuncInv{k}{\cdot}$ is the inverse function of $\KKTFunc{k}{\cdot}$ given by:
	%%%
	%%%%
	\begin{align}
		\KKTFunc{k}{\taskSplit{}}
		\triangleq&\,
		\eCompConst{k}
		\eCompCap{k}
		\workload{}^{\eCompConst{k}}
		\left(\tArrivK{k}\right)^{1-\eCompConst{k}}
%		\hspace{-1mm}
		\taskSplit{}^{\eCompConst{k}-1}
		\nonumber
		\\
		&
		+
		\frac
		{
			\result{}\noise\ln\left(2\right)
		}
		{
			\PathLoss{k}\BW\MRTFuncInv{\targetK{}}
		}
		2^{\frac{\result{}\taskSplit{}}{\BW\left(\tDeparK{k}-\tArrivK{k}\right)}}.
		\label{eq:KKT_inverse function}
	\end{align}
	%%%
	%%%
\end{lemma}
%----------------------------------------------------------------------------
\begin{IEEEproof}
	See Appendix~\ref{app:lem4}.
\end{IEEEproof}
%----------------------------------------------------------------------------

Since $\eCompConst{k}\geq 1$ in \eqref{eq:KKT_inverse function}, the derivative of $\KKTFunc{k}{\taskSplit{}}$ is positive, and therefore, $\KKTFunc{k}{\taskSplit{}}$ is an increasing function.
As $\KKTFunc{k}{\taskSplit{}}$ is strictly increasing with $\taskSplit{}$, $\KKTFuncInv{k}{\KKTeqOpt{}}$ is also strictly increasing with $\KKTeqOpt{}$.
As $\KKTFuncInv{k}{\KKTeqOpt{}}$ is strictly increasing with $\KKTeqOpt{}$,
we can easily obtain $\KKTeqOpt{}$ satisfying $\sum_{k=1}^{\BStotal}\taskSplitOptK{k}=1$ using the bisection method.
Therefore, based on Lemma~\ref{lemma:optimal_x}, we can readily obtain the semi-closed-form optimal solution of Problem~\ref{prob:single_master_data}.

%To provide insights, we consider a homogeneous wireless network consisting of $\BStotal$ identical \acp{RSU}, i.e.,
From Lemma \ref{lemma:optimal_x}, we can obtain the closed-form optimal solution for a homogeneous setup with $\BScovK{k}=\BScovK{}$, 
$\targetK{k}=\targetK{}$, 
$\PathLoss{k}=\PathLoss{}$, 
$\eCompCap{k}=\eCompCap{}$, 
$\eCompConst{k}=\eCompConst{}$, $k\in\BSset$
as follows.
%%%%%%%%%%%%%%%%%%%%%%%%%%%%%%%%%%%%%%%%%%%%%%%%%%%%%%%%%%%%%%%%%%%%%%%%%%%%%
%----------------------------------------------------------------------------
\begin{corollary}[Optimal Solution of Problem \ref{prob:single_master_data} for Homogeneous Setup]
	If 
	$
	\BSfreqM{k}
	\geq
	\frac
	{\workload{}\KKTFuncInv{k}{\KKTeqOpt{}}}
	{\tArrivK{k}}
	$	
	and
	$
	\BSpowerM{k}
	\geq
	\frac
	{\noise}
	{\PathLoss{}\MRTFuncInv{\targetK{}}}
	\left(
	2^{
		\frac
		{\result{}\KKTFuncInv{k}{\KKTeqOpt{}}}
		{\BW\left(\tDeparK{k}-\tArrivK{k}\right)}
	}
	-
	1
	\right)
	$
	,
	the optimal solution is given by:
	%%%
	%%%
	\begin{align}
		\taskSplitOptK{k}
		=
		\max
		\hspace{-0.5mm}
		\left\{
		\KKTFuncTildInv{k}{\KKTeqOpt{}},
		\,\,
		0
		\right\}
		,\quad
		k\in\BSset
		\label{eq:single_opt_data_simple}
	\end{align}
	%%%
	%%%
	where $\KKTeqOpt{}$ satisfies $\sum_{k=1}^{\BStotal}\KKTFuncTildInv{k}{\KKTeqOpt{}}=1$ and 
	$\KKTFuncTildInv{k}{\cdot}$ is the inverse function of $\KKTFuncTild{k}{\cdot}$ given by:
	%%%
	%%%%
	\begin{align}
		\KKTFuncTild{k}{\taskSplit{}}
		\triangleq&
		\eCompConst{}
		\eCompCap{}
		\workload{}^{\eCompConst{}}
		\left(\frac{\velK{}}{\BScovReqSingle+(k-1)\BScovK{}}\right)^{\eCompConst{}-1}
		\taskSplit{}^{\eCompConst{}-1}
		\nonumber
		\\
		&
		+
		\frac
		{
			\result{}\noise\ln\left(2\right)
		}
		{
			\PathLoss{}\BW\MRTFuncInv{\targetK{}}
		}
		2^{\frac{\velK{}\taskSplit{}}{\BW\BScovK{}}}.
		\label{eq:single_KKT_inverse_function_special}
	\end{align}
	%%%
	%%%
	\label{cor:single_opt_data_simple}
\end{corollary}
%----------------------------------------------------------------------------
\begin{IEEEproof}
	First, we prove that $\taskSplitOptK{k}$ is represented by \eqref{eq:single_opt_data_simple}.
	If 
	$
	\BSpowerM{k}
	\geq
	\frac
	{\noise}
	{\PathLoss{}\MRTFuncInv{\targetK{}}}
	\left(
	2^{
		\frac
		{\result{}\KKTFuncInv{k}{\KKTeqOpt{}}}
		{\BW\left(\tDeparK{k}-\tArrivK{k}\right)}
	}
	-
	1
	\right)
	$,
	then
	$
	\min
	\left[
	\frac{\BW\left(\tDeparK{k}-\tArrivK{k}\right)}{\result{}}
	\log_2
	\hspace{-0.5mm}
	\left(
	1
	\hspace{-0.5mm}
	+
	\hspace{-0.5mm}
	\frac
	{\BSpowerM{k}\PathLoss{}\MRTFuncInv{\targetK{}}}
	{\noise}
	\right)
	,
	\max
	\hspace{-0.5mm}
	\left\{
	\KKTFuncInv{k}{\KKTeqOpt{}},
	0
	\right\}
	\right]$
	$=\max
	\hspace{-0.5mm}
	\left\{
	\KKTFuncInv{k}{\KKTeqOpt{}},
	\,\,
	0
	\right\}
	$.
	Moreover, if 
	$
	\BSfreqM{k}
	\geq
	\frac
	{\workload{}\KKTFuncInv{k}{\KKTeqOpt{}}}
	{\tArrivK{k}}
	$,
	then
	$
	\min
	\left[
	\frac{\BSfreqM{k}\tArrivK{k}}{\workload{}}
	,
	\,\,
	\max
	\hspace{-0.5mm}
	\left\{
	\KKTFuncInv{k}{\KKTeqOpt{}},
	\,\,
	0
	\right\}
	\right]
	=
	\max
	\hspace{-0.5mm}
	\left\{
	\KKTFuncInv{k}{\KKTeqOpt{}},
	\,\,
	0
	\right\}
	$.
	Therefore, by \eqref{eq:single_opt_data}, we have \eqref{eq:single_opt_data_simple}.
Therefore, we complete the proof of Corollary~\ref{cor:single_opt_data_simple}.
\end{IEEEproof}
%----------------------------------------------------------------------------
%%%%%%%%%%%%%%%%%%%%%%%%%%%%%%%%%%%%%%%%%%%%%%%%%%%%%%%%%%%%%%%%%%%%%%%%%%%%%
Analogously, by Corollary~\ref{cor:single_opt_data_simple}, we can readily obtain the semi-closed-form optimal solution of Problem~\ref{prob:single_master_data} using the bisection methods.

We can solve $\BStotal$ subproblems by Lemma~\ref{lemma:opt_sol_p4}, each with computational complexity $\mathcal{O}\left(1\right)$.
The master problem can be solved by Lemma~\ref{lemma:optimal_x} with computational complexity $\mathcal{O}\left(\BStotal\right)$.
Therefore, in the single-vehicle case, the overall computational complextiy for solving Problems \ref{prob:single_master_data} and \ref{prob:single_sub_power} is $\mathcal{O}\left(\BStotal\right)$, which is lower than that in the multi-vehicle case.

\section{Online Scenario}
\label{sec:online_case}

In this section, we first present the network model and cooperative offloading model of the edge computing-enabled vehicular network in the online scenario where the information about all vehicles that will enter the network is not available in advance.
Then, we formulate the energy minimization problem in the online scenario and discuss its solution method.

%\/\/\/\/\/\/\/\/\/\/\/\/\/\/\/\/\/\/\/\/\/\/\/\/\/\/\/\/\/\/\/\/\/\/\/\/\/\/\/\/\/\/\/\/\/\/\/\/\/\/\/\/\/\/\/\/\/\/\/\/\/\/\/\/\/\/\/\/\/\/\/\/\/\/\/\/\/\/\/\/\/\/\/\/\/\/\/
%***** x-axis:  [tc][bc][0.7] y-axis: [bc][tc][0.7], legend: [Bl][Bl][0.59]
\begin{figure}[t!]
	\begin{center}
		{
			\includegraphics[width=1.0\columnwidth]{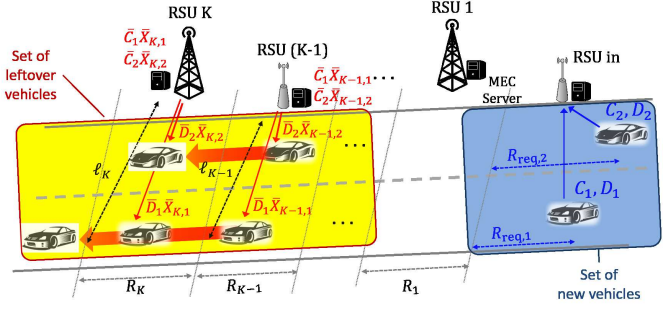}
		}
	\end{center}
	\caption{
		Illustration of the online scenario.
	}
	\label{fig:system_model_online}
\end{figure}
%\/\/\/\/\/\/\/\/\/\/\/\/\/\/\/\/\/\/\/\/\/\/\/\/\/\/\/\/\/\/\/\/\/\/\/\/\/\/\/\/\/\/\/\/\/\/\/\/\/\/\/\/\/\/\/\/\/\/\/\/\/\/\/\/\/\/\/\/\/\/\/\/\/\/\/\/\/\/\/\/\/\/\/\/\/\/\/

\subsection{Network Model and Cooperative Offloading Model in Online Scenario}
\label{sec:online_system}

In the online scenario, we consider the same edge computing-enabled vehicular network consisting of $\BStotal$ multi-antenna \acp{RSU} as in the offline scenario.
The notations for the network parameters are the same as those in the offline scenario.
%Similar to the offline scenario, we consider the edge computing-enabled vehicular network which consists of $\BStotal$ multi-antenna \acp{RSU} in the online scenario.
Unlike the offline scenario, at any time instant $t>0$ when some vehicles, referred to as \emph{new arrivals}, are about to enter the network, some vehicles, called \emph{leftovers}, already exist in the network.
Let $\userTot$ and $\userBarTot$ denote the numbers of new arrivals and leftovers, respectively.
Denote $\userSet\triangleq\{1,\cdots,\userTot\}$ and $\userBarSet\triangleq\left\{\idxBar{1},\cdots,\idxBar{\userTot}\right\}$ as the sets of new arrivals and leftovers, respectively.\footnote{
	Note that in the offline scenario, the set $\userBarSet$ is the empty set and therefore does not have to be considered.
}
For ease of exposition, we assume that the minimum distance between a leftover and a new arrival at any time instant is large enough so that no leftovers and new arrivals will appear in an \ac{RSU}'s coverage area before they leave the network.
In the online scenario, the mobility, task, computation, and communication models of each new arrival and leftover are the same as those of a vehicle in the offline scenario.
Besides, in the online scenario, the notations for the vehicle parameters, task parameters, computation parameters, and communication parameters of each new arrival are identical to those of a vehicle in the offline scenario.
For all $q\in\userBarSet$, let $\velBar{q}$, $\workloadBar{q}$, and $\resultBar{q}$ denote the velocity, the size of the workload, and the size of the computation result for leftover $q$ in the online scenario, respectively.

At time $t$, the task splitting among the $\BStotal$ \acp{RSU} for the leftovers in $\userBarSet$ has already been determined, and the computations and communications for some subtasks have been completed.
We denote $\taskSplitBar{k,q}$ as the fraction of task $q\in\userBarSet$ that has not yet been executed at \ac{RSU} $k\in\BSset$.
Thus, the sizes of the remaining computation workload and the computation result of leftover $q$'s subtask at \ac{RSU} $k$ are $\workloadBar{q}\taskSplitBar{k,q}$ and $\resultBar{q}\taskSplitBar{k,q}$, respectively, where $\taskSplitBar{k,q}\in[0,1]$.
The arrival and departure times of leftover $q$ at \ac{RSU} $k$ are denoted by $\tArrivBar{k,q}$ and $\tDeparBar{k,q}$, respectively.
Let $\userOrderBar{k,n}$ denote the index of the leftover that is the $n$-th one entering \ac{RSU} $k$'s coverage area for all $k\in\BSset$ and $n\in\userBarSet$.
As in the offline scenario, we assume that the controller knows the task parameters of leftovers and new arrivals, i.e., $\workloadBar{k,q}\taskSplitBar{k,q}$, $\resultBar{k,q}\taskSplitBar{k,q}$ and $\workload{k,u}\taskSplit{k,u}$, $\result{k,u}\taskSplit{k,u}$, and the vehicle parameters for leftovers and new arrivals, i.e., $\velBar{q}$, $\tArrivBar{k,q}$, $\tDeparBar{k,q}$ and $\velK{u}$, $\tArrivK{k,u}$, $\tDeparK{k,u}$.

\subsection{Online Optimization}
Let $\taskSplitSet{}\triangleq\left(\taskSplit{k,u}\right)_{k\in\BSset,u\in\userSet}$, $\BSfreqSet{}\triangleq\left(\BSfreqK{k,u}\right)_{k\in\BSset,u\in\userSet}$, $\BSpowerSet{}\triangleq\left(\BSpowerK{k,u}\right)_{k\in\BSset,u\in\userSet}$, $\tCommSet{}\triangleq\left(\tCommK{k,u}\right)_{k\in\BSset,u\in\userSet}$, $\tStartCPSet{}\triangleq\left(\tStartCP{k,u}\right)_{k\in\BSset,u\in\userSet}$, and $\tStartCMSet\triangleq\left(\tStartCM{k,u}\right)_{k\in\BSset,u\in\userSet}$
denote the task splitting, the CPU frequency allocation, the transmission power allocation, the transmission time allocation, the computation starting time allocation, and the transmission starting time allocation for the new arrivals, respectively.
Then, as in the offline scenario, $\taskSplitSet{}$, $\BSfreqSet{}$, $\BSpowerSet{}$, $\tCommSet{}$, $\tStartCPSet{}$, and $\tStartCMSet$ satisfy the constraints in 
\eqref{eq:multi_data_positive},
\eqref{eq:multi_data_sum},
\eqref{eq:multi_freq_positive},
\eqref{eq:multi_freq_max_freq},
\eqref{eq:multi_cp_prev_current_tStart},
\eqref{eq:multi_cp_current_arriv},
\eqref{eq:multi_power_positive},
\eqref{eq:multi_power_max_power},
\eqref{eq:multi_cm_tcm_positive},
\eqref{eq:multi_cm_tStartCM_arriv},
\eqref{eq:multi_cm_current_depart},	
\eqref{eq:multi_cm_prev_current_tStart},
and
\eqref{eq:multi_MISO_MRT_STP_2}.
However, unlike the offline scenario, $\tStartCP{k,u}$ satisfies:
\begin{align}
	\tStartCP{k,u}
	\geq
	t,
	\quad
	k\in\BSset
	,\,\,
	u\in\userSet
\label{eq:online_cp_tStartCP_positive},
\end{align}
rather than \eqref{eq:multi_cp_tStartCP_positive}, since the new arrivals enter the network at time $t$ rather than time 0.

For the leftovers, the task splitting has been completed and cannot be changed, the communication resource allocation at the \acp{RSU} does not have to be updated under the assumption that no leftovers and new arrivals will appear in an \ac{RSU}'s coverage area, and the computation resource allocation at the \acp{RSU} should be updated given the new arrivals at time $t$.
Let $\BSpowerBar{k,q}\in[0,\BSpowerM{k}]$ and $\tCommBar{k,q}\in[0,\tDeparK{k,q}-\tArrivK{k,q}]$ denote the transmission power and the transmission time allocated for transmitting leftover $q$'s subtask $k$ from \ac{RSU} $k$ to leftover $q$, respectively, which have been determined.
Let $\BSfreqBarSet=\left(\BSfreqBar{k,q}\right)_{k\in\BSset,q\in\userBarSet}$ and $\tStartCPBarSet=\left(\tStartCPBar{k,q}\right)_{k\in\BSset,q\in\userBarSet}$ denote the updated CPU frequency allocation and computation starting time allocation for the leftovers, respectively.
As in the offline scenario, we have the following computation constraints for $\BSfreqBarSet$ and $\tStartCPBarSet$:
\begin{align}
	\BSfreqBar{k,q}
	&\geq
	0,
	\quad
	k\in\BSset,
	\,\,
	q\in\userBarSet,
	\label{eq:online_left-freq_positive}
	\\
	\BSfreqBar{k,q}
	&\leq
	\BSfreqM{k},
	\quad
	k\in\BSset,
	\,\,
	q\in\userBarSet,
	\label{eq:online_left-freq_max}
	\\
	\tStartCPBar{k,\userOrderBar{k,n}}
	+
	\frac{\workloadBar{\userOrderBar{k,n}}\taskSplitBar{k,\userOrderBar{k,n}}}{\BSfreqBar{k,\userOrderBar{k,n}}}
	&\leq
	\tStartCPBar{k,\userOrderBar{k,n+1}},
	\nonumber
	\\
	&
	\qquad\quad
	k\in\BSset,
	\,\,
	n\in\userBarSet\backslash\{\userBarTot\},
	\label{eq:online_left-batch_cp_previous}
	\\
	\tStartCPBar{k,\userOrderBar{k,n}}
	+
	\frac{\workloadBar{\userOrderBar{k,n}}\taskSplitBar{k,\userOrderBar{k,n}}}{\BSfreqBar{k,\userOrderBar{k,n}}}
	&\leq
	\tArrivBar{k,\userOrderBar{k,n}},
	\quad
	k\in\BSset,
	\,\,
	n\in\userBarSet.
	\label{eq:online_left-batch_cp_current}
\end{align}

Different from the offline scenario, $\tStartCPBarSet$ satisfies:
\begin{align}
	\tStartCPBar{k,q}
	\geq
	t,
	&\quad
	k\in\BSset,
	\,\,
	q\in\batchSet,
	\label{eq:online_left-tCP_t}
\end{align}
rather than \eqref{eq:multi_cp_tStartCP_positive}.
Furthermore, assuming that no leftovers and new arrivals will appear in an \ac{RSU}'s coverage area, we have the following additional computation constraints for $\tStartCPSet{}$ and $\tStartCPBarSet$:
\begin{align}
	\tStartCPBar{k,\userOrderBar{k,\userBarTot}}
	+
	\frac{\workloadBar{\userOrderBar{k,n}}\taskSplitBar{k,\userOrderBar{k,\userBarTot}}}{\BSfreqBar{k,\userBarTot}}
	&\leq
	\tStartCP{k,\userOrder{k,1}},
	\quad
	k\in\BSset.
	\label{eq:online_left-batch_cp_last_one}
\end{align}

The total energy consumed at the $\BStotal$ \acp{RSU} for serving the $\userTot$ new arrivals, denoted by $\enerTot\left(\taskSplitSet{},\BSfreqSet{},\BSpowerSet{},\tCommSet{}\right)$, is given by \eqref{eq:multi_energy_objective_freq}.
The total energy consumed at the $\BStotal$ \acp{RSU} for serving the $\userBarTot$ leftovers from time $t$, denoted by $\enerTotBar\left(\BSfreqBarSet\right)$, is given by:
%%%
%%%
\begin{align}
	\enerTotBar
	\left(\BSfreqBarSet\right)
	&=
	\sum_{k\in\BSset}
	\sum_{q\in\userBarSet}
	\left(
	\eCompCap{k}
	\workloadBar{q}
	\taskSplitBar{k,q}
	\BSfreqBar{k,q}^{\eCompConst{k}-1}
	+
	\BSpowerBar{k,q}
	\tCommBar{k,q}
	\right).
	\label{eq:online_batch_energy}
\end{align}
%%%
%%%
Therefore, in the online scenario, the total energy consumed at the \acp{RSU} for serving both leftovers and new arrivals from time $t$ is given by:
%%%
%%%
\begin{align}
	\onlineEnerTot
	\left(\taskSplitSet{}, \BSfreqSet{}, \BSpowerSet{}, \tCommSet{}, \BSfreqBarSet\right)
	=&
	\enerTot
	\left(\taskSplitSet{}, \BSfreqSet{}, \BSpowerSet{}, \tCommSet{}\right)
	+
	\enerTotBar
	\left(\BSfreqBarSet\right),
	\label{eq:online_energy_objective}
\end{align}
%%%
%%%
where $\enerTot
	\left(\taskSplitSet{}, \BSfreqSet{}, \BSpowerSet{}, \tCommSet{}\right)$ and $\enerTotBar\left(\BSfreqBarSet\right)$
are given by \eqref{eq:multi_energy_objective_freq} and \eqref{eq:online_batch_energy}, respectively.

In the online scenario, we would like to minimize $\onlineEnerTot
\left(\taskSplitSet{}, \BSfreqSet{}, \BSpowerSet{}, \tCommSet{}, \BSfreqBarSet\right)$ in \eqref{eq:online_energy_objective} by optimizing $\taskSplitSet{}$, $\BSfreqSet{}$, $\BSpowerSet{}$, $\tCommSet{}$, $\tStartCPSet{}$, $\tStartCMSet$, $\BSfreqBarSet$, and $\tStartCPBarSet$ under the constraints in
\eqref{eq:multi_data_positive}\textendash
\eqref{eq:multi_freq_max_freq},	\eqref{eq:multi_cp_prev_current_tStart}\textendash
\eqref{eq:multi_power_max_power},
\eqref{eq:multi_cm_tcm_positive}\textendash
\eqref{eq:multi_cm_prev_current_tStart},
\eqref{eq:multi_MISO_MRT_STP_2},
and
\eqref{eq:online_cp_tStartCP_positive}\textendash
\eqref{eq:online_left-batch_cp_last_one}.
Therefore, the optimization problem is formulated as follows:
%%%%%%%%%%%%%%%%%%%%%%%%%%%%%%%%%%%%%%%%%%%%%%%%%%%%%%%%%%%%%%%%%%%%%%%%%%%%%
%----------------------------------------------------------------------------
\begin{problem}[Energy Minimization in Online Scenario]
	\begin{alignat}{2}
		\onlineEnerOpt
		\triangleq
		\min_{
			\substack{
				\taskSplitSet{}, \BSfreqSet{}, \BSpowerSet{}, \tCommSet{}, \tStartCPSet{}, \tStartCMSet \\
				\BSfreqBarSet, \tStartCPBarSet
			}
		}
%		\hspace{-1mm}
		&  
		\onlineEnerTot
		\left(\taskSplitSet{}, \BSfreqSet{}, \BSpowerSet{}, \tCommSet{}, \BSfreqBarSet\right)
		\nonumber
		%\label{eq:ObjFunc_single_original}
		\\
		\text{s.t.}\qquad
		& 
		% (3)
		\eqref{eq:multi_data_positive}\textendash
		% (4)
%		\eqref{eq:multi_data_sum},
		% (5)
%		\eqref{eq:multi_freq_positive},
		% (6)
		\eqref{eq:multi_freq_max_freq},
		% (9)
		\eqref{eq:multi_cp_prev_current_tStart}\textendash
		% (10)
%		\eqref{eq:multi_cp_current_arriv},
		% (11)
%		\eqref{eq:multi_power_positive},
		% (12)
		\eqref{eq:multi_power_max_power},
		% (16)
		\eqref{eq:multi_cm_tcm_positive}\textendash
		% (17)
%		\eqref{eq:multi_cm_tStartCM_arriv},
		% (18)
%		\eqref{eq:multi_cm_current_depart},
		% (19)
		\eqref{eq:multi_cm_prev_current_tStart},
		% (27)
		\eqref{eq:multi_MISO_MRT_STP_2},
		\nonumber
		\\
		&
		% (50)
		\eqref{eq:online_cp_tStartCP_positive}\textendash
%		\eqref{eq:online_left-freq_positive},
%		\eqref{eq:online_left-freq_max},
%		\eqref{eq:online_left-batch_cp_previous},
%		\eqref{eq:online_left-batch_cp_current},
%		\eqref{eq:online_left-tCP_t},
		\eqref{eq:online_left-batch_cp_last_one}.
		\nonumber
	\end{alignat}
	\label{prob:online_energy_min_freq}
\end{problem}
%----------------------------------------------------------------------------
%%%%%%%%%%%%%%%%%%%%%%%%%%%%%%%%%%%%%%%%%%%%%%%%%%%%%%%%%%%%%%%%%%%%%%%%%%%%%

Given the network parameters and the parameters of both new arrivals and leftovers, the controller can handle this optimization.
The structure of Problem~\ref{prob:online_energy_min_freq} for the online scenario is similar to the structure of Problem~\ref{prob:multi_energy_min_freq} for the offline scenario,
as problems have identical constraints, i.e., \eqref{eq:multi_data_positive}\textendash	\eqref{eq:multi_freq_max_freq},
\eqref{eq:multi_cp_prev_current_tStart}\textendash
\eqref{eq:multi_power_max_power},
\eqref{eq:multi_cm_tcm_positive}\textendash
\eqref{eq:multi_cm_prev_current_tStart},
\eqref{eq:multi_MISO_MRT_STP_2},
and the structures of
\eqref{eq:online_cp_tStartCP_positive}\textendash
\eqref{eq:online_left-batch_cp_last_one} are very similar to those of
\eqref{eq:multi_freq_positive}, \eqref{eq:multi_freq_max_freq}, \eqref{eq:multi_cp_tStartCP_positive}\textendash\eqref{eq:multi_cp_current_arriv}.
Therefore, Problem \ref{prob:online_energy_min_freq} can be solved using the proposed methods for the offline scenario.
We omit the details due to the page limitation.

\section{Numerical Results for Offline Scenario}\label{sec:numerical}

In this section, we numerically show the total energy consumptions of our proposed solutions for the single-vehicle and the multi-vehicle cases in the offline scenario.
Unless otherwise specified, we set $\BStotal=20$, $\noise=-80$dBm, $B=5$MHz, $\PL=4$, $\kappa_{k}=10^{-11}$, $\CPUenergy{k}=3$, $k\in\BSset$, and $\workload{u}=1000\result{u}$, and $\targetK{u}=0.95$, $u\in\userSet$.
In the single-vehicle case, we set $\BScovReq{1}=300$m, $\result{1}=300$MB, and $\velK{1}=75$km/h.
In the multi-vehicle case, we set $\userTot=2$, $\BScovReq{1}=300$m, $\BScovReq{2}=400$m, $\result{1}=\left(\result{}-\frac{1}{2}\delta_D\right)$MB, $\result{2}=\left(\result{}+\frac{1}{2}\delta_D\right)$MB,  $\velK{1}=\left(v-\frac{1}{2}\delta_v\right)$km/h, and $\velK{2}=\left(v+\frac{1}{2}\delta_v\right)$km/h, where $\result{}=50$MB, $\delta_D=0$MB, $v=80$km/h, and $\delta_v=10$km/h, unless otherwise specified.
It is worth noting that $\result{}$ and $\velK{}$ represent the average computation result size and velocity of the two vehicles, respectively, and $\delta_D$ and $\delta_v$ represent the differences in the computation result sizes and velocities of the two vehicles, respectively.
We consider two network setups in each case, i.e., the single-tier network and the two-tier network.
In the single-tier network, we set $\BScovK{k}=500$m, $\BSpowerM{k}=50$dBm,  $\BSfreqM{k}=1.1$GHz, $k\in\BSset$.
In the two-tier network, we set
$\BScovK{k}=600$m,
$\BSpowerM{k}=55$dBm, $\BSfreqM{k}=1.2$GHz, $k\in\{1,3,\cdots,19\}$ and
$\BScovK{k}=400$m, $\BSpowerM{k}=45$dBm, $\BSfreqM{k}=1.0$GHz, $k\in\{2,4,\cdots,20\}$.
Note that both network setups have identical $\sum_{k\in\BSset}\BSpowerM{k}$ and $\sum_{k\in\BSset}\BSfreqM{k}$ for a fair comparison.

\subsection{Single-Vehicle Case}

To assess the total energy consumption of our proposed solution in the single-vehicle case, we consider two baseline schemes, namely, the \ac{BEF} scheme and the \ac{BEL} scheme.
From the constraints in \eqref{eq:single_opt_freq_max_freq} and \eqref{eq:single_opt_tcm_max_power_feasibility} of Problem~\ref{prob:single_master_data}, we have $\taskSplit{k,1}\leq\taskSplitMax{k,1}$, $k\in\BSset$, where
$\taskSplitMax{k,1}
\triangleq
\min
\left\{
\frac{\BSfreqM{k}\tArrivK{k,1}}{\workload{1}}, \,\,
\frac{\BW\left(\tDeparK{k,1}
	%			\hspace{-0.8mm}
	-
	%			\hspace{-0.5mm}
	\tArrivK{k,1}\right)}{\result{1}}
\log_2
\left(
%		\hspace{-0.5mm}
1+
\frac
{\BSpowerM{k}\PathLoss{k}\MRTFuncInv{\targetK{}}}
{\noise}
%		\hspace{-0.5mm}
\right)
\right\}$ represents the maximum fraction of the task that can be handled by \ac{RSU} $k$ under the simulation setup.
The \ac{BEF} scheme splits the task according to  $\taskSplit{k,1}=\taskSplitMax{k,1}$, $k\in\{1,\cdots,\bar{k}-1\}$, $\taskSplit{\bar{k},1}=1-\sum_{i=1}^{\bar{k}}\taskSplitMax{i,1}$, and $\taskSplit{k,1}=0$, $k\in\{\bar{k}+1,\cdots,\BStotal\}$, where $\bar{k}\triangleq\min\left\{k\in\BSset \, \lvert\,\sum_{i=1}^{\BStotal}\taskSplitMax{i,1}>1\right\}$.
On the other hand, the \ac{BEL} scheme splits the task according to $\taskSplit{k,1}=\taskSplitMax{k,1}$, $k\in\{\underline{k}+1,\cdots,\BStotal\}$, $\taskSplit{\underline{k}}=1-\sum_{i=\underline{k}+1}^{\BStotal}\taskSplitMax{i,1}$, and $\taskSplit{k,1}=0$, $k\in\{1,\cdots,\underline{k}-1\}$,
where $\underline{k}\triangleq\max\left\{k\in\BSset \, \lvert\,\sum_{i=1}^{\BStotal}\taskSplitMax{i,1}>1\right\}$.

%\/\/\/\/\/\/\/\/\/\/\/\/\/\/\/\/\/\/\/\/\/\/\/\/\/\/\/\/\/\/\/\/\/\/\/\/\/\/\/\/\/\/\/\/\/\/\/\/\/\/\/\/\/\/\/\/\/\/\/\/\/\/\/\/\/\/\/\/\/\/\/\/\/\/\/\/\/\/\/\/\/\/\/\/\/\/\/
\begin{figure}[t!]
	\centering
	\subfigure[Velocity of the vehicle.]
	{
		\psfrag{enerTot}[bc][tc][0.8] {Total Energy Consumption [J]}
		\psfrag{velocity}[Bc][bc][0.8] {Vehicle's velocity, $\velK{1}$ [km/h]}
		\psfrag{ProposedSingleTier}[Bl][Bl][0.6] {$\,$Proposed (Single)}
		\psfrag{BEFsingle}[Bl][Bl][0.6] {$\,$BEF (Single)}
		\psfrag{BELsingle}[Bl][Bl][0.6] {$\,$BEL (Single)}
		\psfrag{ProposedTwoTier}[Bl][Bl][0.6] {$\,$Proposed (Two)}
		\psfrag{BEFTwo}[Bl][Bl][0.6] {$\,$BEF (Two)}
		\psfrag{BELTwo}[Bl][Bl][0.6] {$\,$BEL (Two)}
		\includegraphics[width=0.45\textwidth]{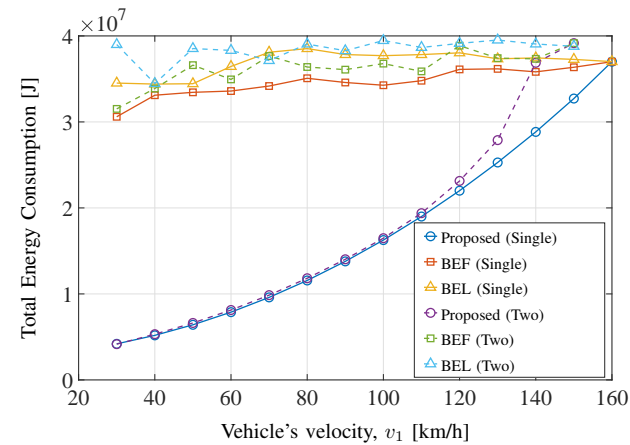}
		\label{subfig:single_vel}
	}
	\subfigure[Size of the computation result.]
	{
		\psfrag{enerTot}[bc][tc][0.8] {Total Energy Consumption [J]}
		\psfrag{reqDataMB}[Bc][bc][0.8] {Computation Result Size, $\result{1}$ [MB]}
		\psfrag{ProposedSingleTier}[Bl][Bl][0.6] {$\,$Proposed (Single)}
		\psfrag{BEFsingle}[Bl][Bl][0.6] {$\,$BEF (Single)}
		\psfrag{BELsingle}[Bl][Bl][0.6] {$\,$BEL (Single)}
		\psfrag{ProposedTwoTier}[Bl][Bl][0.6] {$\,$Proposed (Two)}
		\psfrag{BEFTwo}[Bl][Bl][0.6] {$\,$BEF (Two)}
		\psfrag{BELTwo}[Bl][Bl][0.6] {$\,$BEL (Two)}
		\includegraphics[width=0.45\textwidth]{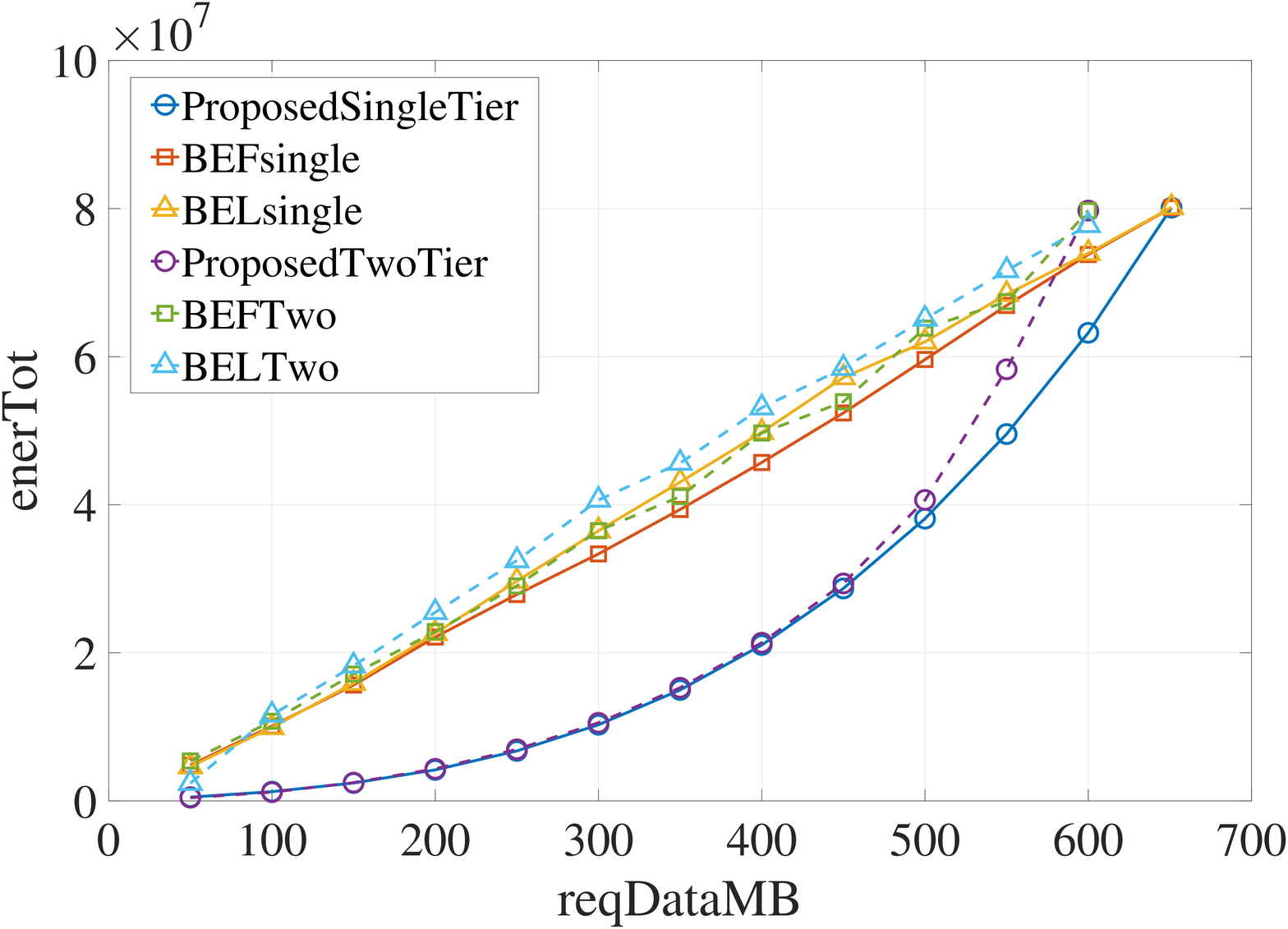}
		\label{subfig:single_result_size}
	}
	\caption{
	Total energy consumption versus the velocity of the vehicle $\velK{1}$ and the computation result size of the vehicle $\result{1}$ in the single-tier and two-tier network.
	}
	\label{fig:single_result_vel}
\end{figure}
%\/\/\/\/\/\/\/\/\/\/\/\/\/\/\/\/\/\/\/\/\/\/\/\/\/\/\/\/\/\/\/\/\/\/\/\/\/\/\/\/\/\/\/\/\/\/\/\/\/\/\/\/\/\/\/\/\/\/\/\/\/\/\/\/\/\/\/\/\/\/\/\/\/\/\/\/\/\/\/\/\/\/\/\/\/\/\/

Fig.~\ref{subfig:single_vel} and \ref{subfig:single_result_size} illustrate the total energy consumption versus the vehicle's velocity $\velK{1}$ and the size of the computation result $\result{1}$, respectively.
Note that for the single-tier network, the problem is not feasible when $\result{1}=300$MB, $\velK{1}>150$km/h or $\result{1}>650$MB, $\velK{1}=75$km/h.
Moreover, for the two-tier network, the problem is not feasible when $\result{1}=300$MB, $\velK{1}>160$km/h or $\result{1}>600$MB, $\velK{1}=75$km/h.
From Fig.~\ref{subfig:single_vel}, we can see that the total energy consumption of the proposed solution 
decreases as $\velK{1}$ decreases, whereas the total energy consumption of each baseline scheme does not show a noticeable trend with $\velK{1}$.
This is because the time for serving the vehicle's task increases as $\velK{1}$ decreases, but the baseline schemes do not effectively adapt to $\velK{1}$ as the proposed solution.
From Fig.~\ref{subfig:single_result_size}, we can see that the total energy consumption of each scheme increases with $\result{1}$.
This is because as $\result{1}$ increases, each \ac{RSU} allocates more CPU frequency and transmission power to execute and transmit the subtasks, respectively.
Additionally, from Fig.~\ref{fig:single_result_vel}, we can see that our proposed solution outperforms both baseline schemes in each network setup.
This is because the proposed solution optimally splits the vehicle's task and allocates resources at the \acp{RSU} according to the network parameters in the single-vehicle case.
Moreover, we can see that the total energy consumption of each scheme in the two-tier network is larger than that in the single-tier network,
indicating that heterogeneity influences resource utilization efficiency.

%\/\/\/\/\/\/\/\/\/\/\/\/\/\/\/\/\/\/\/\/\/\/\/\/\/\/\/\/\/\/\/\/\/\/\/\/\/\/\/\/\/\/\/\/\/\/\/\/\/\/\/\/\/\/\/\/\/\/\/\/\/\/\/\/\/\/\/\/\/\/\/\/\/\/\/\/\/\/\/\/\/\/\/\/\/\/\/
\begin{figure}[t!]
	\centering
	\subfigure[Average velocity of the two vehicles.]
	{
		\psfrag{enerTot}[bc][tc][0.8] {Total Energy Consumption [J]}
		\psfrag{avgV}[Bc][bc][0.8] {Average Velocity, $v$ [km/h]}
		\psfrag{ProposedSingleTier}[Bl][Bl][0.6] {$\,$Proposed (Single)}
		\psfrag{BEFsingle}[Bl][Bl][0.6] {$\,$BEF (Single)}
		\psfrag{BELsingle}[Bl][Bl][0.6] {$\,$BEL (Single)}
		\psfrag{ProposedTwoTier}[Bl][Bl][0.6] {$\,$Proposed (Two)}
		\psfrag{BEFtwo}[Bl][Bl][0.6] {$\,$BEF (Two)}
		\psfrag{BELtwo}[Bl][Bl][0.6] {$\,$BEL (Two)}
		\includegraphics[width=0.45\textwidth]{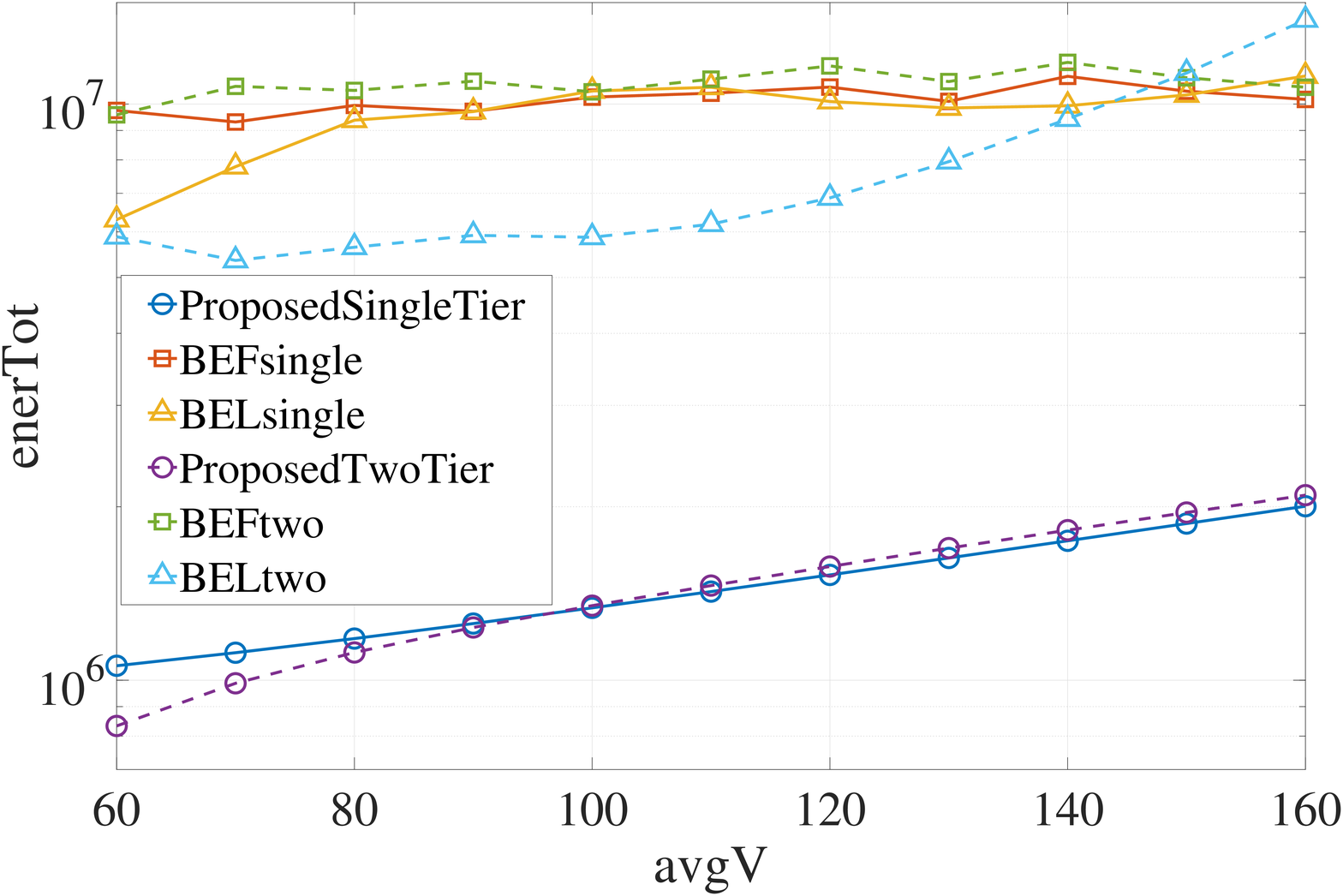}
		\label{subfig:multi_avgV}
	}
	\subfigure[Velocity difference between the two vehicles.]
	{
		\psfrag{enerTot}[bc][tc][0.8] {Total Energy Consumption [J]}
		\psfrag{diffV}[Bc][bc][0.8] {Velocity Difference, $\delta_v$ [km/h]}
		\psfrag{ProposedSingleTier}[Bl][Bl][0.6] {$\,$Proposed (Single)}
		\psfrag{BEFsingle}[Bl][Bl][0.6] {$\,$BEF (Single)}
		\psfrag{BELsingle}[Bl][Bl][0.6] {$\,$BEL (Single)}
		\psfrag{ProposedTwoTier}[Bl][Bl][0.6] {$\,$Proposed (Two)}
		\psfrag{BEFtwo}[Bl][Bl][0.6] {$\,$BEF (Two)}
		\psfrag{BELtwo}[Bl][Bl][0.6] {$\,$BEL (Two)}
		\includegraphics[width=0.45\textwidth]{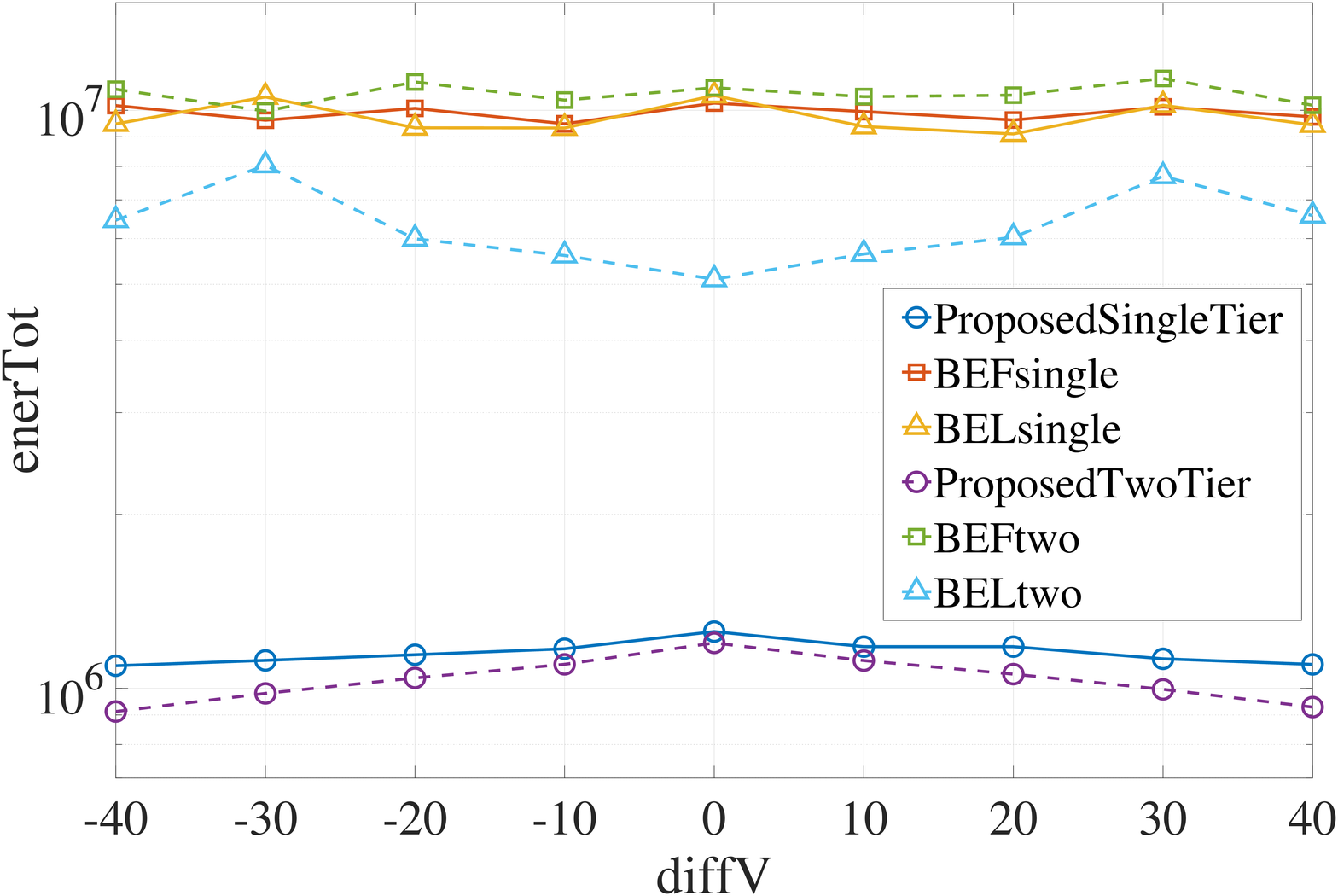}
		\label{subfig:multi_diffV}
	}
	\caption{
		Total energy consumption versus the average velocity $v$ and the velocity difference $\delta_v$ of two vehicles in the single-tier and two-tier network.
	}
\label{fig:multi_velocity}
\end{figure}
%\/\/\/\/\/\/\/\/\/\/\/\/\/\/\/\/\/\/\/\/\/\/\/\/\/\/\/\/\/\/\/\/\/\/\/\/\/\/\/\/\/\/\/\/\/\/\/\/\/\/\/\/\/\/\/\/\/\/\/\/\/\/\/\/\/\/\/\/\/\/\/\/\/\/\/\/\/\/\/\/\/\/\/\/\/\/\/

%\/\/\/\/\/\/\/\/\/\/\/\/\/\/\/\/\/\/\/\/\/\/\/\/\/\/\/\/\/\/\/\/\/\/\/\/\/\/\/\/\/\/\/\/\/\/\/\/\/\/\/\/\/\/\/\/\/\/\/\/\/\/\/\/\/\/\/\/\/\/\/\/\/\/\/\/\/\/\/\/\/\/\/\/\/\/\/
\begin{figure}[t!]
	\centering
	\subfigure[Average computation result size of the two vehicles.]
	{
		\psfrag{enerTot}[bc][tc][0.8] {Total Energy Consumption [J]}
		\psfrag{avgD}[Bc][bc][0.8] {Average Computation Result Size, $\result{}$ [MB]}
		\psfrag{ProposedSingleTier}[Bl][Bl][0.6] {$\,$Proposed (Single)}
		\psfrag{BEFsingle}[Bl][Bl][0.6] {$\,$BEF (Single)}
		\psfrag{BELsingle}[Bl][Bl][0.6] {$\,$BEL (Single)}
		\psfrag{ProposedTwoTier}[Bl][Bl][0.6] {$\,$Proposed (Two)}
		\psfrag{BEFtwo}[Bl][Bl][0.6] {$\,$BEF (Two)}
		\psfrag{BELtwo}[Bl][Bl][0.6] {$\,$BEL (Two)}
		\includegraphics[width=0.45\textwidth]{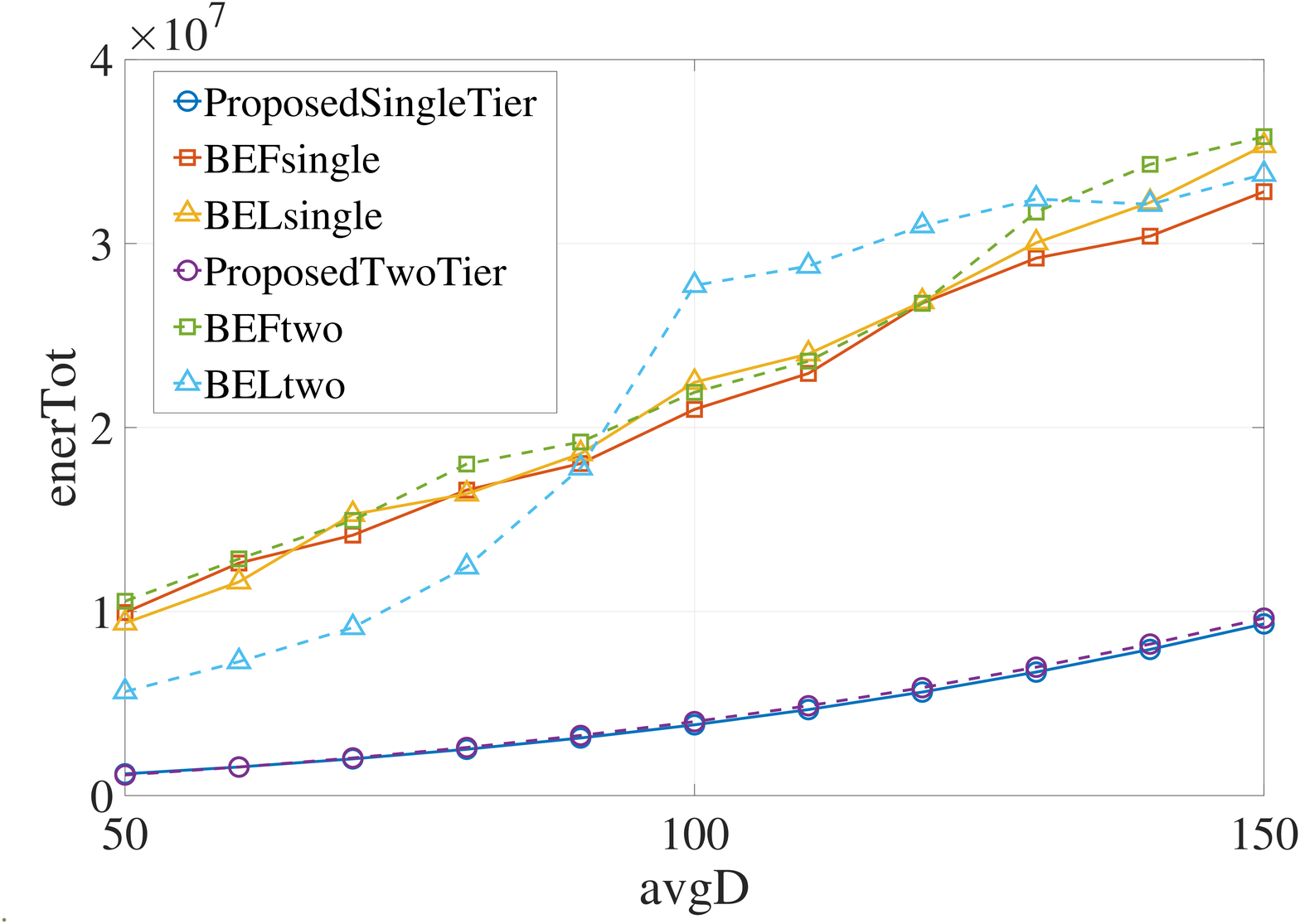}
		\label{subfig:multi_avgResult}
	}
	\subfigure[Computation result size difference between the two vehicles.]
	{
		\psfrag{enerTot}[bc][tc][0.8] {Total Energy Consumption [J]}
		\psfrag{diffD}[Bc][bc][0.8] {Computation Result Size Difference, $\delta_D$ [MB]}
		\psfrag{ProposedSingleTier}[Bl][Bl][0.6] {$\,$Proposed (Single)}
		\psfrag{BEFsingle}[Bl][Bl][0.6] {$\,$BEF (Single)}
		\psfrag{BELsingle}[Bl][Bl][0.6] {$\,$BEL (Single)}
		\psfrag{ProposedTwoTier}[Bl][Bl][0.6] {$\,$Proposed (Two)}
		\psfrag{BEFtwo}[Bl][Bl][0.6] {$\,$BEF (Two)}
		\psfrag{BELtwo}[Bl][Bl][0.6] {$\,$BEL (Two)}
		\includegraphics[width=0.45\textwidth]{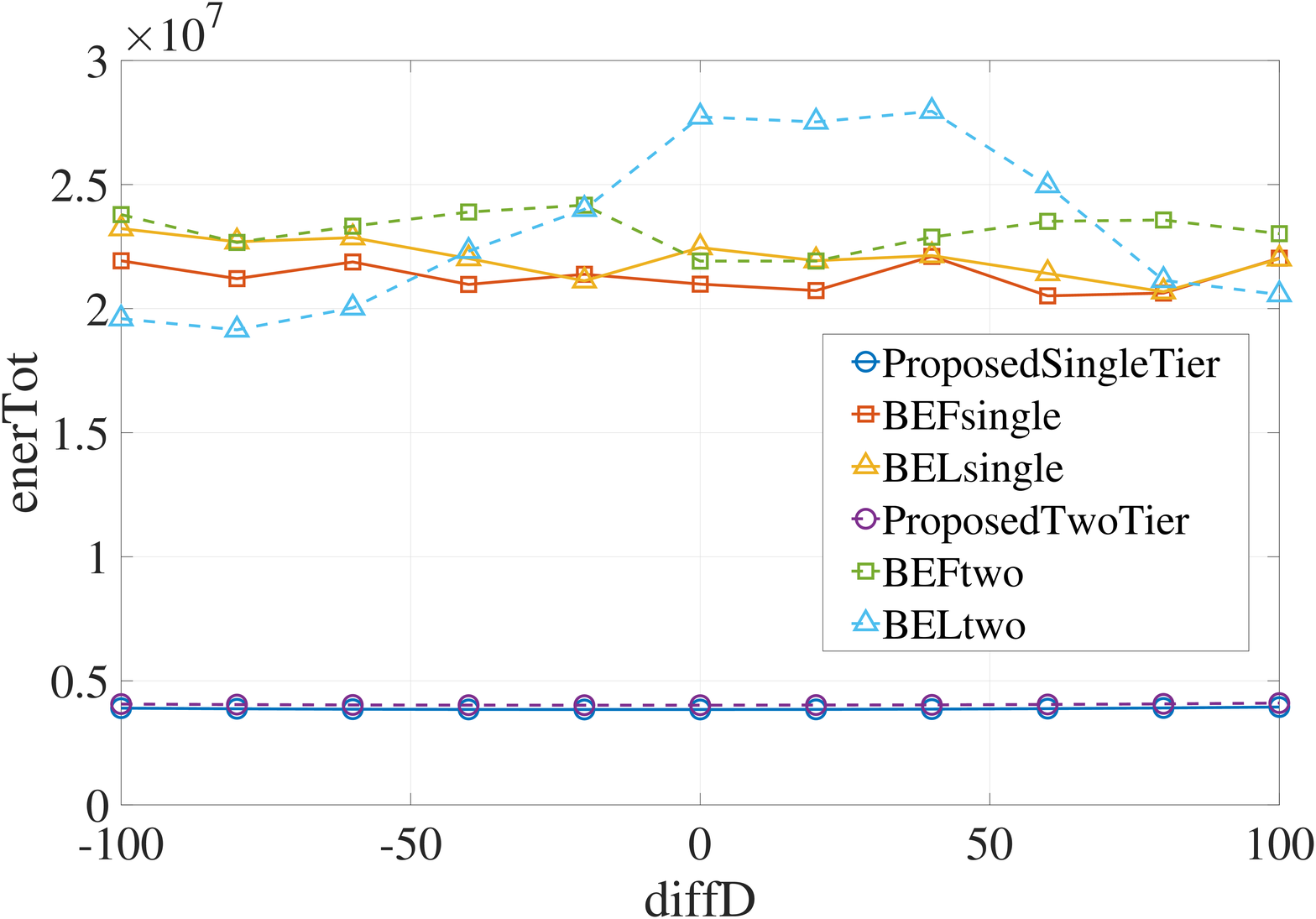}
		\label{subfig:multi_diffResult}
	}
	\caption{
		Total energy consumption versus the average computation result size $\result{}$ and the computation result size difference $\delta_D$ of two vehicles in the single-tier and two-tier network.
	}
	\label{fig:multi_result}
\end{figure}
%\/\/\/\/\/\/\/\/\/\/\/\/\/\/\/\/\/\/\/\/\/\/\/\/\/\/\/\/\/\/\/\/\/\/\/\/\/\/\/\/\/\/\/\/\/\/\/\/\/\/\/\/\/\/\/\/\/\/\/\/\/\/\/\/\/\/\/\/\/\/\/\/\/\/\/\/\/\/\/\/\/\/\/\/\/\/\/

\subsection{Multi-Vehicle Case}
In the multi-vehicle case, we consider two baseline schemes that are generalized from those in the single-vehicle case and are also termed the \ac{BEF} scheme and the \ac{BEL} scheme for ease of exposition.
%As there exists two vehicles driving with different velocities,
Each \ac{RSU} serves the vehicles in their arriving order in both baseline schemes as in the proposed solution.
Specifically, from the constraints in \eqref{eq:single_opt_freq_max_freq} and \eqref{eq:single_opt_tcm_max_power_feasibility} of Problem~\ref{prob:single_master_data}, 	
we have $\taskSplit{k,\userOrder{k,1}}\leq\taskSplitMax{k,\userOrder{k,1}}$, $k\in\BSset$ for vehicle $\userOrder{k,1}$, where
	$\taskSplitMax{k,\userOrder{k,1}}
	\triangleq
	\min
	\left\{
	\frac{\BSfreqM{k}\tArrivK{k,\userOrder{k,1}}}{\workload{\userOrder{k,1}}}, \,\,
	\frac{\BW\left(\tDeparK{\userOrder{k,1}}
		%			\hspace{-0.8mm}
		-
		%			\hspace{-0.5mm}
		\tArrivK{\userOrder{k,1}}\right)}{\result{\userOrder{k,1}}}
	\log_2
	\left(
	%		\hspace{-0.5mm}
	1
	+
	\right.
	\right.$
	$
	\left.
	\left.
	\frac
	{\BSpowerM{k}\PathLoss{k}\MRTFuncInv{\targetK{}}}
	{\noise}
	%		\hspace{-0.5mm}
	\right)
	\right\}$
represents the maximum fraction of vehicle $\userOrder{k,1}$'s task that can be handled by \ac{RSU} $k$ under the simulation setup.
On the other hand, from the constraints in \eqref{eq:multi_cp_prev_current_tStart}, \eqref{eq:multi_cp_current_arriv}, \eqref{eq:multi_cm_current_depart}, and \eqref{eq:multi_cm_prev_current_tStart} of Problem~\ref{prob:multi_energy_min_freq},
we have $\taskSplit{k}\leq\taskSplitMax{k,\userOrder{k,2}}$, $k\in\BSset$ for vehicle $\userOrder{k,2}$, where
$
\taskSplitMax{k,\userOrder{k,2}}
\triangleq
\min
\left\{
\frac
{\BSfreqM{k}\left(\tArrivK{k,\userOrder{k,2}}-\tArrivK{k,\userOrder{k,1}}\right)}
{\workload{\userOrder{k,2}}}, \,\,
\right.$
$\left.
\frac
{
	\max
	\left\{
	0,
	\BW\left(
	\tDeparK{k,\userOrder{k,2}}
	-
	\max\left(\tArrivK{k,\userOrder{k,2}},\tDeparK{k,\userOrder{k,1}}\right)
	\right)
	\right\}
	\log_2
	\left(
	%		\hspace{-0.5mm}
	1+
	\frac
	{\BSpowerM{k}\PathLoss{k}\MRTFuncInv{\targetK{}}}
	{\noise}
	%		\hspace{-0.5mm}
	\right)
}
{\result{1}}
\right\}.
$
Based on $\taskSplitMax{k,\userOrder{k,1}}$ and $\taskSplitMax{k,\userOrder{k,2}}$ defined above, the \ac{BEF} scheme and the \ac{BEL} scheme split the tasks as in the multi-vehicle case.

Fig. \ref{subfig:multi_avgV} and \ref{subfig:multi_diffV} show the total energy consumption versus the average velocity of the two vehicles, $v$, and the velocity difference of the two vehicles, $\delta_v$, respectively.
Analogous to Fig. \ref{subfig:multi_avgV}, we can see that the total energy consumption of the proposed solution decreases as $\velK{}$ decreases, whereas the total energy consumption of each baseline scheme does not show an evident trend with $\velK{}$.
Moreover, from Fig. \ref{subfig:multi_diffV}, we can see that the total energy consumption of the proposed solution decreases with $|\delta_v|$, whereas the total energy consumption of each baseline scheme does not show an apparent trend with $|\delta_v|$.
This is because the time for serving the first arriving vehicle at each \ac{RSU} increases with $|\delta_v|$, but the baseline schemes do not effectively adapt to $|\delta_v|$ as the proposed solution.
Fig. \ref{subfig:multi_avgResult} and \ref{subfig:multi_diffResult} show the total energy consumption versus the average computation result size of the two vehicles, $\result{}$, and the computation result size difference between the two vehicles, $\delta_D$, respectively.
Analogously, from Fig. \ref{subfig:multi_avgResult}, we can see that the total energy consumption of each scheme increases with $\result{}$, and
%This is because as $\result{}$ increases, each \ac{RSU} allocates more CPU frequency and transmission power to execute and transmit the subtasks of two vehicles.
from Fig. \ref{subfig:multi_diffResult}, we can see that the total energy consumption of our proposed solution is less dependent on $\delta_D$ compared to the baseline schemes.
%In other workds, our proposed solution successfully optimizes the task splitting and the resource allocation by effectively adopting to the network parameters.
Also, Fig.~\ref{fig:multi_velocity} and Fig.~\ref{fig:multi_result} show that our proposed solution, which represents the optimal offloading design, outperforms the baseline schemes in the single-tier and two-tier networks.
%This is because the proposed solution optimally splits the vehicles' tasks and allocates resources at the \acp{RSU} according to the network parameters in the multi-vehicle case.

%---------------------------------------------------------------------------%
%                            Sec: Conclusion                                %
%---------------------------------------------------------------------------%

\section{Conclusion}\label{sec:conclusion}

This paper develops the energy-efficient cooperative offloading scheme for edge computing-enabled vehicular networks.
Specifically, we first establish the cooperative offloading model for the offline scenario and online scenario, which splits the task into multiple subtasks and offloads them to different \acp{RSU} located ahead along the route of the vehicle.
Then, for each scenario, we formulate the total energy consumption minimization problem with respect to the task splitting ration, computation resource, and communication resource.
The formulated problem is a challenging non-convex problem with a considerably large number of variables and constraints.
In the offline scenario, we show that the original optimization problem can be transformed into a convex problem with fewer variables and constraints, and obtain the optimal solutions in both multi-vehicle and single-vehicle cases.
In the online scenario, we show that the underlying problem structure is similar to the one for the offline scenario and hence can be solved using the same method.
Finally, numerical results show that the proposed solutions achieve a significant gain in the total energy consumption over all baseline schemes.
We show that the total energy consumption of the proposed solutions decreases with the velocity difference of vehicles and increases with the vehicle's velocity and the size of the computation results.
We also show that there exist maximum vehicle's velocity and maximum computation result size that the vehicle can be successfully served.

%---------------------------------------------------------------------------%
%                                Appendix                                   %
%---------------------------------------------------------------------------%

\begin{appendix}
	
\subsection{Proof of Theorem~\ref{theorem:off-multi_convex_transform}}\label{app:theo_multi_convex_transform}

Since \eqref{eq:multi_data_positive}, \eqref{eq:multi_cm_tcm_positive}, and \eqref{eq:multi_MISO_MRT_STP_2} imply \eqref{eq:multi_power_positive}, we can ignore \eqref{eq:multi_power_positive}.
Lemma \ref{lemma:multi_opt_power} implies that \eqref{eq:multi_MISO_MRT_STP_2} holds with equality at the optimal solution. Thus, we can replace $\BSpowerK{k,u}$ in Problem~\ref{prob:multi_energy_min_freq} with
$
\frac{\noise}{\PathLoss{k}\MRTFuncInv{\targetK{u}}}
\left(
2^{\frac{\result{u}\taskSplit{k,u}}{\BW\tCommK{k,u}}}
-
1
\right)
$,
for all $k\in\BSset$, $u\in\userSet$, and ignore the constraint in \eqref{eq:multi_MISO_MRT_STP_2}.
Furthermore, by substituting $\BSpowerK{k,u}=
	\frac{\noise}{\PathLoss{k}\MRTFuncInv{\targetK{u}}}
	\left(
	2^{\frac{\result{u}\taskSplit{k,u}}{\BW\tCommK{k,u}}}
	-
	1
	\right)$
into \eqref{eq:multi_power_max_power}, we have \eqref{eq:multi_MRT_optimal_power}.
Since \eqref{eq:multi_data_positive} and \eqref{eq:multi_MRT_optimal_power} imply \eqref{eq:multi_cm_tcm_positive}, we can ignore \eqref{eq:multi_cm_tcm_positive}.
Consequently, Problem \ref{prob:multi_energy_min_freq} can be equivalently transformed to the following problem.

	%%%%%%%%%%%%%%%%%%%%%%%%%%%%%%%%%%%%%%%%%%%%%%%%%%%%%%%%%%%%%%%%%%%%%%%%%%%%%
	%----------------------------------------------------------------------------
	\begin{problem}[Equivalent Problem of Problem \ref{prob:multi_energy_min_freq} In Multi-Vehicle Case]
		\begin{alignat}{2}
			\min_{
				\substack{
					\taskSplitSet{}, \BSfreqSet{},\tCommSet{},\\
					\tStartCPSet{}, \tStartCMSet
				}
				%			\taskSplitSet{}, \BSfreqSet{}, \tCommSet{}, \tStartCPSet{}, \tStartCMSet
			}
			\,\,
			&  
			\sum_{k=1}^{\BStotal}
			\sum_{u=1}^{\userTot}
			\left(
			\enerK{\CP,k,u}
			\left(\taskSplit{k,u},\BSfreqK{k,u}\right)
			+
			\enerKTild{\CM,k,u}
			\left(\taskSplit{k,u}, \tCommK{k,u}\right)
			\right)
			\nonumber
			%\label{eq:ObjFunc_single_original}
			\\
			\text{s.t.}\quad
			& 
			% (3)
			\eqref{eq:multi_data_positive},
			% (4)
			\eqref{eq:multi_data_sum},
			% (5)
			\eqref{eq:multi_freq_positive},
			% (6)
			\eqref{eq:multi_freq_max_freq},
			% (8)
			\eqref{eq:multi_cp_tStartCP_positive},
			% (9)
			\eqref{eq:multi_cp_prev_current_tStart},
			% (10)
			\eqref{eq:multi_cp_current_arriv},
			\nonumber
			\\
			&
			% (15)
			\eqref{eq:multi_cm_tStartCM_arriv},
			% (16)
			\eqref{eq:multi_cm_current_depart},
			% (17)
			\eqref{eq:multi_cm_prev_current_tStart},
			\eqref{eq:multi_MRT_optimal_power}.
			\nonumber
		\end{alignat}
		\label{prob:multi_energy_min_freq_opt_power}
	\end{problem}
	%----------------------------------------------------------------------------
	%%%%%%%%%%%%%%%%%%%%%%%%%%%%%%%%%%%%%%%%%%%%%%%%%%%%%%%%%%%%%%%%%%%%%%%%%%%%%
	where $\enerKTild{\CM,k,u}\left(\taskSplit{k,u}, \tCommK{k,u}\right)$ is obtained from \eqref{eq:multi_energy_cm_x_tcm}.

By substituting $\BSfreqK{k,u}=\frac{\workload{u}\taskSplit{k,u}}{\tCompK{k,u}}$ into the objective function of Problem~\ref{prob:multi_energy_min_freq_opt_power}, we have \eqref{eq:multi_energy_objective_x_tcp_tcm}.
By replacing $\BSfreqK{k,u}$ in \eqref{eq:multi_freq_max_freq}, \eqref{eq:multi_cp_prev_current_tStart}, and \eqref{eq:multi_cp_current_arriv} with $\frac{\workload{u}\taskSplit{k,u}}{\tCompK{k,u}}$, we have \eqref{eq:multi_cp_tcp_max_freq}, \eqref{eq:multi_cp_tcp_prev_current_tStart}, and \eqref{eq:multi_cp_tcp_current_arriv}, respectively.
Since \eqref{eq:multi_data_positive} and \eqref{eq:multi_cp_tcp_max_freq} imply \eqref{eq:multi_freq_positive}, we can ignore \eqref{eq:multi_freq_positive}.
%Since $\frac{\workload{u}\taskSplit{k,u}}{\tCompK{k,u}}$ is non-negative, we can ignore $\frac{\workload{u}\taskSplit{k,u}}{\tCompK{k,u}}\geq 0$, which is obtained from \eqref{eq:multi_freq_positive}.
Therefore, Problem~\ref{prob:multi_energy_min_freq_opt_power} is equivalent to Problem~\ref{prob:multi_energy_min_tcp}.

\subsection{Proof of Theorem~\ref{theorem:single_equivalence}}
\label{app:theo_single_equiv}

First, we prove that $\tStartCPOptK{k}=0$ and $\tStartCMOptK{k}=\tArrivK{k}$.
For any $\tStartCPSet{}$ and $\tStartCMSet$ satisfying \eqref{eq:multi_cp_tStartCP_positive} and \eqref{eq:multi_cm_tStartCM_arriv}, respectively, the feasible set of the other variables $\left(\taskSplitSet{}, \BSfreqSet{}, \BSpowerSet{}, \tCommSet{}\right)$ is given by:
\begin{align}
	\mathcal{Z}\left(\tStartCPSet{},\tStartCMSet\right)
	\triangleq&
	\left\{
	\left(\taskSplitSet{}, \BSfreqSet{}, \BSpowerSet{}, \tCommSet{}\right)
	\lvert\,
	% x=0
	\eqref{eq:multi_data_positive},
	% \sum x=1
	\eqref{eq:multi_data_sum},
	% f>=0
	\eqref{eq:multi_freq_positive},
	% f<=F
	\eqref{eq:multi_freq_max_freq},
	\right.
	\nonumber
	\\
	&
	\left.
	\eqref{eq:multi_cp_current_arriv},
	% p>=0
	\eqref{eq:multi_power_positive},
	% p<=P
	\eqref{eq:multi_power_max_power},
	% t_cm>=0
	\eqref{eq:multi_cm_tcm_positive},
	\eqref{eq:multi_cm_current_depart},	
	\eqref{eq:multi_MISO_MRT_STP_2}
	\right\}.
\end{align}
Note that the constraints in \eqref{eq:multi_cp_prev_current_tStart} and \eqref{eq:multi_cm_prev_current_tStart} in Problem~\ref{prob:multi_energy_min_freq} are void in the single-vehicle case.
Then, we can express Problem~\ref{prob:multi_energy_min_freq} in the following equivalent form.
%%%%%%%%%%%%%%%%%%%%%%%%%%%%%%%%%%%%%%%%%%%%%%%%%%%%%%%%%%%%%%%%%%%%%%%%%%%%%
%----------------------------------------------------------------------------
\begin{problem}[Equivalent Problem of Problem~\ref{prob:multi_energy_min_freq}]
	\begin{alignat}{2}
		\enerOpt
		\triangleq
		\min_{\taskSplitSet{}, \BSfreqSet{}, \BSpowerSet{}, \tCommSet{}, \tStartCPSet{}, \tStartCMSet}\quad
		&  
		\enerTot
		\left(\taskSplitSet{}, \BSfreqSet{}, \BSpowerSet{}, \tCommSet{}\right)
		\nonumber
		%\label{eq:ObjFunc_single_original}
		\\
		\text{s.t.}\quad\,\,\,
		& 
		% s_cp>=0
		\eqref{eq:multi_cp_tStartCP_positive},
		% s_cm>T_arrv
		\eqref{eq:multi_cm_tStartCM_arriv},
		\nonumber
		\\
		&
		\left(\taskSplitSet{}, \BSfreqSet{}, \BSpowerSet{}, \tCommSet{}\right)
		\in
		\mathcal{Z}\left(\tStartCPSet{},\tStartCMSet\right).
		\nonumber
	\end{alignat}
	\label{prob:single_scm_scp-optimal}
\end{problem}
%----------------------------------------------------------------------------
%%%%%%%%%%%%%%%%%%%%%%%%%%%%%%%%%%%%%%%%%%%%%%%%%%%%%%%%%%%%%%%%%%%%%%%%%%%%%
Note that $\tStartCPSet{}$ and $\tStartCMSet$ affect $\mathcal{Z}\left(\tStartCPSet{},\tStartCMSet\right)$ only through the constraints in \eqref{eq:multi_cp_current_arriv} and \eqref{eq:multi_cm_current_depart}.
From \eqref{eq:multi_cp_current_arriv} and \eqref{eq:multi_cm_current_depart}, $\mathcal{Z}\left(\tStartCPSet{},\tStartCMSet\right)$ achieves the largest when $\tStartCP{k}=0$ and $\tStartCM{k}=\tArrivK{k}$.
Therefore, we can prove that $\tStartCPOptK{k}=0$ and $\tStartCMOptK{k}=\tArrivK{k}$ are optimal.

Next, we show that Problem~\ref{prob:multi_energy_min_freq} is equivalent to Problem~\ref{prob:single_energy_min}.
For given $\tStartCPOptK{k}=0$ and $\tStartCMOptK{k}=\tArrivK{k}$, it remains to minimize $\enerTot
\left(\taskSplitSet{}, \BSfreqSet{}, \BSpowerSet{}, \tCommSet{}\right)$ with respect to  $\left(\taskSplitSet{},\BSfreqSet{},\BSpowerSet{},\tCommSet{}\right)$.
Moreover, \eqref{eq:multi_cp_current_arriv} and \eqref{eq:multi_cm_current_depart} become \eqref{eq:single_cp_arriv} and 
\eqref{eq:single_cm_dep_arriv}, respectively.
Therefore, we can prove that the optimal solutions of Problem~\ref{prob:multi_energy_min_freq} and Problem~\ref{prob:single_energy_min} satisfy \eqref{eq:single_equiv_sol}.

\subsection{Proof of Theorem~\ref{theorem:single_p456_equiv}}\label{app:theo2}

First, we characterize optimality properties of Problem \ref{prob:single_energy_min}.
For given $\taskSplit{k}$ and $\BSpowerK{k}$, the objective function increases with $\tCommK{k}$ and $\BSfreqK{k}$, for all $k\in\BSset$.
Besides, the constraints in \eqref{eq:multi_data_positive}, \eqref{eq:multi_data_sum}, \eqref{eq:multi_power_positive}, and \eqref{eq:multi_power_max_power} are not related to $\tCommK{k}$, $\BSfreqK{k}$, $k\in\BSset$.
Thus, by contradiction, we can show that the inequality constraints in \eqref{eq:multi_MISO_MRT_STP_2} and \eqref{eq:single_cp_arriv} are active at $\left(\taskSplitOptSet, \BSfreqOptSet, \BSpowerKoptSet, \tCommOptSet\right)$.
Therefore, a solution of Problem~\ref{prob:single_energy_min} satisfies:
	\begin{align}
		\tCommOptK{k}
		&=
		\frac
		{\result{}\taskSplitOptK{k}}
		{\BW\log_2 \left(1+\frac{\BSpowerOptK{k}\PathLoss{k}}{\noise}\MRTFuncInv{\targetK{}}\right)}
		,\quad k\in\BSset,
		\label{eq:single_opt_freq}
		\\
		\BSfreqOptK{k}
		&=
		\frac
		{
			\workload{}\taskSplitOptK{k}
		}
		{
			\tArrivK{k}
		}
		,\quad k\in\BSset.
		\label{eq:single_opt_tcm}
	\end{align}
Thus, without loss of optimality, we can replace $\tCommK{k}$ and $\BSfreqK{k}$ in Problem \ref{prob:single_energy_min} with
$
\frac
{\result{}\taskSplit{k}}
{\BW\log_2 \left(1+\frac{\BSpowerK{k}\PathLoss{k}}{\noise}G^{-1}\left(\targetK{}\right)\right)}
$
and
$
\frac
{
	\workload{}\taskSplit{k}
}
{
	\tArrivK{k}
}
$
respectively, for all $k\in\BSset$, $u\in\userSet$, and ignore the constraints in \eqref{eq:multi_MISO_MRT_STP_2} and \eqref{eq:single_cp_arriv}.
Moreover, by substituting $\BSfreqK{k}=\frac
{
	\workload{}\taskSplit{k}
}
{
	\tArrivK{k}
}$ and $\tCommK{k}=\frac
{\result{}\taskSplit{k}}
{\BW\log_2 \left(1+\frac{\BSpowerK{k}\PathLoss{k}}{\noise}G^{-1}\left(\targetK{}\right)\right)}$ into \eqref{eq:multi_freq_max_freq} and \eqref{eq:single_cm_dep_arriv},
we have \eqref{eq:single_opt_freq_max_freq} and \eqref{eq:single_opt_tcm_rate_constraint}.
For Problem~\ref{prob:single_energy_min} to be feasible, the lower bound of $\BSpowerK{k}$ in \eqref{eq:single_opt_tcm_rate_constraint} must be smaller than the upper bound in  \eqref{eq:multi_power_max_power}, and thus, we have \eqref{eq:single_opt_tcm_max_power_feasibility}.
Since \eqref{eq:multi_data_positive} and \eqref{eq:single_opt_tcm_rate_constraint} imply \eqref{eq:multi_power_positive}, we ignore \eqref{eq:multi_power_positive}.
Moreover, \eqref{eq:multi_data_positive} implies \eqref{eq:multi_freq_positive} and \eqref{eq:multi_cm_tcm_positive}, and therefore, we can ignore \eqref{eq:multi_freq_positive} and \eqref{eq:multi_cm_tcm_positive}.
Therefore, Problem~\ref{prob:single_energy_min} can be equivalently transformed to the following problem.
\begin{problem}[Equivalent Problem of Problem~\ref{prob:single_energy_min}]
	\begin{alignat}{2}
		\min_{\taskSplitSet, \BSpowerSet{}}\quad
		&  
		\sum_{k=1}^{\BStotal}
		\left(
		\eCompCap{k}
		\workload{}^{\eCompConst{k}}
		\left(\tArrivK{k}\right)^{1-\eCompConst{k}}
		\taskSplit{k}^{\eCompConst{k}}
		\right.
		\nonumber
		\\
		&
		\left.
		+
		\frac
		{\result{}\taskSplit{k}\BSpowerK{k}}
		{\BW\log_2 \left(1+\frac{\BSpowerK{k}\PathLoss{k}}{\noise}\MRTFuncInv{\targetK{}}\right)}
		\right)
		\nonumber
		%\label{eq:ObjFunc_single_original}
		\\
		\text{s.t.}\quad\,\,\,
		& 
		% (3)
		\eqref{eq:multi_data_positive},
		% (4)
		\eqref{eq:multi_data_sum},
		% (12)
		\eqref{eq:multi_power_max_power},
		% (40)
		\eqref{eq:single_opt_freq_max_freq},
		% (41)
		\eqref{eq:single_opt_tcm_max_power_feasibility},
		% (43)
		\eqref{eq:single_opt_tcm_rate_constraint}.
		\nonumber
		%\label{eq:Const1_single_original}
	\end{alignat}
	\label{prob:single_energy_min-equiv}
\end{problem}

The constraints in \eqref{eq:multi_data_positive}, \eqref{eq:multi_data_sum}, \eqref{eq:single_opt_freq_max_freq}, and \eqref{eq:single_opt_tcm_max_power_feasibility} are only related to $\taskSplit{k}$, while the constraint in \eqref{eq:multi_power_max_power} is only related to $\BSpowerK{k}$.
Thus, we can equivalently convert Problem~\ref{prob:single_energy_min-equiv} to a master problem in Problem~\ref{prob:single_master_data}, which optimizes $\taskSplitSet{}$, and $\BStotal$ subproblems in Problem~\ref{prob:single_sub_power}, which optimize $\BSpowerK{k}$, $k\in\BSset$, respectively.
%In Problem~\ref{prob:single_master_data}, \eqref{eq:single_opt_tcm_max_power_feasibility} is the feasibility condition obtained from the constraints in \eqref{eq:multi_power_max_power} and \eqref{eq:single_opt_tcm_rate_constraint}.
Therefore, we complete the proof of Theorem \ref{theorem:single_p456_equiv}.

\subsection{Proof of Lemma~\ref{lemma:optimal_x}}\label{app:lem4}

First, by relaxing the coupling constraint in \eqref{eq:multi_data_sum}, we obtain the partial Lagrange function:
%%%
%%%
\begin{align}
	L
	\left(
	\taskSplitSet{},\KKTeq{}
	\right)
	=&
	\KKTeq{}
	\left(
	1
	-
	\sum_{k=1}^{\BStotal}
	\taskSplit{k}
	\right)
	+
	\sum_{k=1}^{\BStotal}
	\left\{
	\eCompCap{k}
	\workload{}^{\eCompConst{k}}
	\left(\tArrivK{k}\right)^{1-\eCompConst{k}}
	\taskSplit{k}^{\eCompConst{k}}
	%		\right.
	%		\nonumber
	%		\\
	%		&
	%		\left.
	\right.
	\\
	&
	\left.
	+
	\frac
	{\noise\left(\tDeparK{k}-\tArrivK{k}\right)}
	{\PathLoss{k}\MRTFuncInv{\targetK{}}}
	\left(
	2^{\frac{\compResultK{k}}{\BW\left(\tDeparK{k}-\tArrivK{k}\right)}}
	-
	1
	\right)
	\right\},
	\label{eq:single_lagrange}
\end{align}
%%%
%%%
Now we obtain the \ac{KKT} conditions:
%%%
%%%
\begin{align}
	&\eqref{eq:multi_data_positive},\,\,
	\eqref{eq:single_opt_freq_max_freq},\,\,
	\eqref{eq:single_opt_tcm_max_power_feasibility},\,\,	
	1
	-
	\sum_{k=1}^{\BStotal}
	\taskSplit{k}
	=
	0,
	\nonumber
	\\
	&\frac
	{
		\partial
		L
		\left(
		\taskSplitSet{},\KKTeq{}
		\right)
	}
	{\partial \taskSplit{k}}	
	=
	\eCompConst{k}
	\eCompCap{k}
	\workload{}^{\eCompConst{k}}
	\left(\tArrivK{k}\right)^{1-\eCompConst{k}}
	\taskSplit{k}^{\eCompConst{k}-1}
			\nonumber
			\\
	&+
	\frac
	{
		\result{}\noise\ln\left(2\right)
	}
	{
		\PathLoss{k}\BW\MRTFuncInv{\targetK{}}
	}
	2^{\frac{\result{}\taskSplit{k}}{\BW\left(\tDeparK{k}-\tArrivK{k}\right)}}
	-
	\KKTeq{}
	=
	0,
	\quad
	k\in\BSset.
	\label{eq:single_kkt_lagra_zero}
\end{align}
%%%
%%%
From \eqref{eq:single_kkt_lagra_zero}, we obtain
$
\taskSplit{k}
=
\KKTFuncInv{k}{\KKTeq{}},
$
where $\KKTFunc{k}{\cdot}$ is given by \eqref{eq:KKT_inverse function}.
Thus, we have
%%%
%%%
\begin{align}
	\taskSplitOptK{k}
	=&
	\max
	\hspace{-0.5mm}
	\left[
	\hspace{-0.5mm}
	\min
	\hspace{-0.5mm}
	\left\{
	\,
	\hspace{-0.5mm}
	\frac{\BW\left(\tDeparK{k}
		\hspace{-0.8mm}
		-
		\hspace{-0.5mm}
		\tArrivK{k}\right)}{\result{}}
	\log_2
	\left(
	1+
	\frac
	{\BSpowerM{k}\PathLoss{k}\MRTFuncInv{\targetK{}}}
	{\noise}
	\right)
	,
			\right.
			\right.
			\nonumber
			\\
			&\qquad\,\,
			\left.
			\left.
	\quad
	\frac{\BSfreqM{k}\tArrivK{k}}{\workload{}}
	,
	\,\,\,
	\KKTFuncInv{k}{\KKTeqOpt{}}
	\right\}
	,\,\,
	0
	\right],
	\,\,
	k\in\BSset.
	\label{eq:single_opt_data_max_min}
\end{align}
%%%
%%%
It is obvious that 
$
\frac{\BW\left(\tDeparK{k}
	\hspace{-0.8mm}
	-
	\hspace{-0.5mm}
	\tArrivK{k}\right)}{\result{}}
\log_2
\left(
1+
\frac
{\BSpowerM{k}\PathLoss{k}\MRTFuncInv{\targetK{}}}
{\noise}
\right)
$
and
$
\frac{\BSfreqM{k}\tArrivK{k}}{\workload{}}
$
%the first and second terms of the min function
in \eqref{eq:single_opt_data_max_min} 
are non-negative values.
Thus, we can represent \eqref{eq:single_opt_data_max_min} by \eqref{eq:single_opt_data}.
Therefore, we complete the proof of Lemma \ref{lemma:optimal_x}.

\end{appendix}

\bibliographystyle{IEEEtran}

\bibliography{IEEEabrv,StringDefinitions,bib_EC-VN}

\end{document}